\documentclass[11pt,a4paper]{article}
\pdfoutput=1
\usepackage[english]{babel}
\usepackage[latin1]{inputenc}
\usepackage[T1]{fontenc}
\usepackage{amsfonts,amsbsy,bm,euscript,mathrsfs}
\usepackage{amssymb,faktor,slashed}
\usepackage{color}
\usepackage[tbtags]{amsmath}
\usepackage[bookmarks=true,colorlinks=true,linkcolor=black,citecolor=black,urlcolor=black,bookmarksnumbered,linktocpage=true,hyperfigures=true,linktoc=all]{hyperref}
\usepackage{graphicx}
\usepackage{chngcntr}
\usepackage{makecell}
\usepackage{mathtools}

\usepackage{afterpage}

\usepackage{enumerate}

\usepackage[a4paper,text={160mm,247mm},centering]{geometry}

\numberwithin{equation}{section}

\makeatletter
\renewcommand\section{\@startsection {section}{1}{\z@}%
{-3.5ex \@plus -1ex \@minus -.2ex}%
{2.3ex \@plus.2ex}%
{\normalfont\large\bfseries}}
\renewcommand\subsection{\@startsection{subsection}{2}{\z@}%
{-3.25ex\@plus -1ex \@minus -.2ex}%
{1.5ex \@plus .2ex}%
{\normalfont\normalsize\bfseries}}
\makeatother

\expandafter\def\expandafter\bfseries\expandafter{\bfseries\ifmmode\else\boldmath\fi}
\expandafter\def\expandafter\mdseries\expandafter{\mdseries\ifmmode\else\unboldmath\fi}
\expandafter\def\expandafter\normalfont\expandafter{\normalfont\ifmmode\else\unboldmath\fi}

\providecommand{\href}[2]{#2}
\newcommand{\arxivlink}[1]{\href{http://arxiv.org/abs/#1}{[arXiv:#1]}}
\newcommand{\doilink}[2]{\href{http://doi.org/#2}{#1}}

\newcommand{\mathsym}[1]{{}}
\def\id{\protect{{1 \kern-.28em{\rm l}}}}
\def\be{\begin{eqnarray}}
\def\ee{\end{eqnarray}}

\def\tr{{\rm tr}}
\def\ha{\tfrac{1}{2}}

\def\a{\alpha}
\def\b{\beta}

\def\g{\gamma}

\def\det{\hbox{det}}

\def\Tr{{\rm Tr}}

\def\l {\lambda}

\def\O{{\mathcal O}}
\def\m{\mu}

\def\foot{\footnote}
\newcommand{\rf}[1]{(\ref{#1})}

\def\no{\nonumber}

\def\la{\label}
\def\l{\lambda}

\def\p{\phi}

\def\varpi{{\rm w}}

\def\del{\partial}
\def\s{\sigma}

\def\ed{\end{document}}
\newcommand{\mc}{\mathcal }
\def\iffa{\iffalse}

\def\ad{{\rm ad}}
\def\d{\delta}

\def\L{\mathcal{L} }

\def\sms{$\sigma$-models}

\def\Ad{\text{Ad}}

\def\Lie{\operatorname{Lie}}

 \def \h {{\rm h}}
\def\dt{\tfrac{d}{dt}}
\def\ddt{\frac{d}{dt}}

\def\ka{{\kappa}}

\def\ed{\end{document}}

\def\cg{c\,}

\def\e{\varepsilon}

\def\g{\gamma}

\def\sms{$\s$-models\ }

\def \foot{\footnote}\def \l {\lambda}\def \iffa {\iffalse}
 \def \a  {\alpha} \def \ha {{1\ov 2}}
\def \ed {\end{document}}

\def \la {\label}

\def \tr {{\rm tr}}
\def \ha {{1 \over 2}}

\def \cg {{\rm c}}

\def \ccg {{\cg_{_G}}}

\def\cG{\ccg}

\def \ha {\tfrac{1}{2}}

\newtheorem{BCA}{Bianchi Completeness Assumption}[section]
\usepackage{float}
\newcommand\PP{{\mathcal P}}
\newcommand\tet{\tilde \eta }

\begin{document}

\ 

\vspace{0.5cm}

\vspace{2.5cm}

\begin{center}

{\Large\bf Universal 1-loop divergences for
 integrable sigma models}

\vspace{1.5cm}
{
Nat Levine$^{a,b}$
}

\vspace{0.8cm}

{
\em \vspace{0.15cm}
$^{a}$Laboratoire de Physique,  \qquad \quad  $^b$Institut Philippe Meyer,\\
\vspace{0.15cm}
 {\'E}cole Normale Sup{\'e}rieure,\\
\vspace{0.15cm}
Universit{\'e} PSL, CNRS, Sorbonne Universit{\'e}, Universit{\'e} Paris Cit{\'e}, \\
\vspace{0.15cm}
24 rue Lhomond, F-75005 Paris, France

}

\vspace{1.2cm}{
\begingroup\ttfamily\small
nat.levine @ phys.ens.fr
\endgroup}

\end{center}

\vspace{1.5cm}

\begin{abstract}
We present a simple, new method for the 1-loop renormalization of integrable $\sigma$-models. By treating equations of motion and Bianchi identities on an equal footing, we derive `universal' formulae for the 1-loop on-shell divergences, generalizing case-by-case computations in the literature. Given a choice of poles for the classical Lax connection, the divergences take a theory-independent form in terms of the Lax currents (the residues of the poles), assuming a `completeness' condition on the zero-curvature equations. We compute these divergences for a large class of theories with simple poles in the Lax connection. We also show that $\mathbb{Z}_T$ coset models of `pure-spinor' type and their recently constructed $\eta$- and $\lambda$-deformations are 1-loop renormalizable, and 1-loop scale-invariant when the Killing form vanishes.
\end{abstract}
  
\newpage 
\tableofcontents

\setcounter{footnote}{0}
\setcounter{section}{0}
\begin{center}
{\large \bf
 }
\end{center}

\section{Introduction}
It is natural to believe that, in the absence of quantum anomalies, the structures associated with integrability (higher symmetries, etc.)\ will be preserved under renormalization. This paper is concerned with a particular semi-classical incarnation of this idea: that 
 the space of classically integrable theories is stable under the leading 1-loop RG flow.   
 
One may say that a 2d field theory is classically integrable if its classical equations of motion are equivalent to the flatness of a 1-parameter family of Lax connections,\foot{We denote by $\xi^a$ ($a=0,1$) the co-ordinates of 2d Minkowski space with signature $\eta_{ab} = \text{diag}(-+)$. We will use  the standard 2d light-cone co-ordinates $\xi^\pm = \ha(\xi^0 \pm \xi^1)$. The antisymmetric 2d Levi-Civita tensor $\e_{ab}$ is defined with $\e_{01}=1$. The target space indices are denoted $m,n=1,\ldots,D$. The Lax connection will always be valued in a (super)algebra $\Lie(G)$, with generators labelled by $\a,\b,\g$  (see appendix \ref{Apeq} for further conventions). The complex spectral parameter is denoted $z$.}$^,$\foot{Formally for classical integrability one should also demand a sufficient number of commuting charges. This can be ensured by assuming a certain form for the Poisson brackets of the Lax (see, e.g., \cite{maillet}).}
 \be
 [\del_+ + L_+(z), \del_- + L_-(z) ] = 0 \ , \qquad \qquad L_\pm \in \Lie(G) \ . \la{zc}
 \ee
  In the context of 2d $\s$-models,
\be \begin{aligned}
 \L &= - (G_{mn}(x) \, \eta^{ab} + B_{mn}(x)\, \e^{ab} ) \,  \del_a x^m \, \del_b x^n \ \  ,  \ \ \   \qquad   S = \frac{1}{4\pi\a'} \int d^2\xi \ \L   \ , \la{sm}\\
 &= (G(x) + B(x))_{mn} \,  \del_+ x^m \, \del_- x^n  
 \end{aligned} \ee
this condition selects an interesting subspace of $\s$-model couplings, or target space geometries $(G, B)$; an interesting goal is to understand and classify this subspace.

 The 1-loop RG flow of general \sms \rf{sm} is a `generalized Ricci flow' in the space of geometries $(G,B)$ \cite{friedan},\foot{We denote $t=\log\mu$ where $\mu$ is a renormalization scale, and will denote by $\Lambda$ a UV momentum cutoff. Here and in the rest of the paper, we set $\a'=1$.}
 \be
 \frac{d}{dt} (G+B)_{mn} = R_{mn} -\tfrac{1}{4} H_{mpq} H_n{}^{pq} - \ha \nabla^p H_{pmn} \ . \la{beta}
 \ee
We expect the subspace of classically integrable couplings to be stable, or closed, under this flow (see, e.g., \cite{intRG} for background on this idea). Equivalently, if a theory's classical equations of motion admit a flat Lax connection, then so should its 1-loop `effective equations of motion' (those obtained by varying the 1-loop effective action).

Indeed, there is much evidence for this conjecture: integrable $\s$-models with a few couplings are found to be renormalizable at 1-loop, with just those couplings running and the existence of a flat Lax pair being preserved. Such examples include the $\eta$-deformed \cite{Squellari} and $\l$-deformed \cite{Itsios:2014lca} models, and the integrable $G^N$ \cite{DLSS} and $G\times G/H$ \cite{LT} models. Moreover, we are not aware of any counterexamples.

\bigskip

In this paper we shall take a first step towards proving the renormalizability of integrable models, by computing universal formulae for their 1-loop divergences in terms of the currents appearing in the Lax connection. The key idea  is that, since integrability forces the equations of motion to take a universal zero-curvature form
  \rf{zc}, then the 1-loop RG flow will also be universal, as it is determined by the linearized equations of motion.  The only model-dependent information will come from (i) the analytic structure of the spectral (i.e. $z$) dependence of the Lax connection and (ii) the relation between the currents in the Lax connection and the  fields.

For example, for integrable models whose Lax connections have the  analytic structure
\be
L_\pm(z) = \frac{1}{1\pm z} \mc A_\pm \ , \la{Li}
 \ee
we will show that the on-shell 1-loop divergences take
 the 
 universal form\foot{We denote the 1-loop on-shell effective action as $\widehat S^{(1)} = \frac{1}{4\pi} \int d^2 \xi \, \widehat \L^{(1)}$. The symbol $\cG$ denotes the dual Coxeter number of $G$ (see appendix \ref{Apeq}).}
\be
\ddt \widehat \L^{(1)} = -\ha \cG \,  \Tr[ \mc A_+ \mc A_- ] \ . \la{for}
\ee
Here the current $\mc A_\pm$ may take any form in terms of the fields of the model, up to some mild assumptions that we discuss below.
As we will show below, the formula  \rf{for} matches and generalizes the known results for the 1-loop $\b$-functions of the principal chiral model (PCM) with and without Wess-Zumino (WZ) term,  its $\eta$- and $\l$-deformations\foot{The formula \rf{for} previously appeared in the particular case of the $\l$-model in \cite{ah}, where it was also obtained by observing that the equations of motion are the same as the PCM up to a redefinition of the current $\mc A_\pm$. Our approach is different in that: (i) we will consider a general on-shell background field, which is technically required to claim renormalizability (as opposed to computing $\b$-functions \textit{assuming} renoramlizability); (ii) we will justify the treatment of equations of motion and Bianchi identities on an equal footing by a path integral argument; and (iii) our approach is much more general, applying not just for the $\l$-model but for a broad class of integrable theories.}, its non-abelian T-dual (NATD) and its pseudo-dual,  with $\mc A_\pm$ taken to be the corresponding Lax current in each case.

We expect the universality of 1-loop divergences to be a generic feature, with a single formula like \rf{for} applying for each choice of poles of the Lax connection.
Indeed, we will compute these `universal' formulae for 
a large class of theories with only simple poles in the Lax connection. This class includes, but is not limited to,  theories obtained from affine Gaudin models, such as integrable $G^N$ models \cite{DLMV} --- and indeed our results match the known 1-loop $\b$-functions for $G\times G$ models \cite{DLSS,LT}.

We will also focus on  the Lax connection corresponding to $\mathbb{Z}_T$ coset models of `pure-spinor' type \cite{Berkovits,Young}.
Again we derive a single formula for the 1-loop on-shell divergences of any theory with a Lax connection of this same analytic structure (assuming the eligiblity conditions below). In particular, as well as the undeformed pure-spinor $\mathbb{Z}_T$ cosets,  our result  applies to their $\eta$- and $\l$-deformations that were recently constructed in \cite{Hoare}.
As a result, we will show that these models  are
1-loop renormalizable and compute their $\b$-functions (confirming the conjecture of \cite{Hoare}). 
In the particular case of supercosets with vanishing Killing form, we show that all theories of this class are 1-loop scale-invariant.

\bigskip
We will give a general path integral argument that such a universal formula for 1-loop divergences applies for classically integrable models with a generic choice of poles in the Lax connection.  Noting that, typically, some of the zero-curvature equations \rf{zc} will be Bianchi identities (satisfied off-shell), we 
  make a crucial assumption that these Bianchi identities are `complete'. That is, they are sufficient to re-write the Lax currents in terms of the physical fields, at least to leading order in a background field expansion.

We will use this assumption to show that the 1-loop on-shell divergences are insensitive to the distinction between equations of motion and Bianchi identities.
  This fact is intuitive because Bianchi identities and equations of motion play the same role 
 in correlation functions up to $\hbar$-suppressed contact terms,
\be \la{ward}
\langle \text{Bianchi} \cdots \rangle = 0  \ , \qquad \quad  \langle \text{EOM} \cdots \rangle = \hbar \, (\text{contact} )\ ,
\ee
so one should expect them to be interchangeable at the leading 1-loop order. As a result, different theories realizing the same Lax connection are equivalent at the level of 1-loop divergences, 
since they differ only by a designation of Bianchi identities and equations of motion.


In this paper we use a background field method and assume all background fields to be on-shell, so we are computing on-shell divergences. In the $\s$-model context, that means we drop possible diffeomorphism contributions  $L_V (G(x)+B(x))_{mn}$ to the $\beta$-function \rf{beta} (where $L_V$ is the Lie derivative along a vector $V$). This is advantageous in the sense of removing the  ambiguity corresponding to field renormalizations, so it is actually easier to detect certain properties, e.g.\ renormalizability. However, it is disadvantageous in the context of string $\s$-models, where 2d Weyl invariance demands that these invisible off-shell terms take a particular form
 \mbox{$V_m = \del_m \phi + \O(\a')$}
 in terms of the dilaton background $\phi$. We will return to this point in section \ref{out}.\sloppypar

\bigskip
This paper is structured as follows. 
In section \ref{PCS}  we focus on models with Lax connection of the same structure as the principal chiral model (eq.\ \rf{Li} above).
 Section \ref{gen} contains the general path integral argument that underlies our approach.
 
Section \ref{CD} focusses on models related to `pure-spinor' $\mathbb{Z}_T$ cosets. Section \ref{gsim} discusses a general class of theories with only simple poles in the Lax connection. Section \ref{Disc}  is a summary and discussion of the results, and section \ref{out} considers the future directions.

In appendix \ref{A} we derive sufficient conditions for the Bianchi `completeness' assumption and show that they are satisfied for all models considered in the paper. In appendix \ref{Apeq} we state our conventions for (super)algebras and derive an identity in terms of structure constants. 
In appendix \ref{T} we 
show that the 1-loop equivalence between equations of motion and Bianchi identities
is consistent with the standard fact that T-duality commutes with the 1-loop RG flow.

\section{Principal chiral model and relatives \la{PCS}}
First let us consider a simple analytic structure for the Lax connection,
\be
 L_\pm(z) = \frac{1}{1\pm z} \mc A_\pm \ . \la{fc} 
\ee
Here $z$ is a complex spectral parameter. 

One example of a theory with such a Lax connection is the principal chiral model (PCM),\foot{Here and below, $\Tr$ is taken to be a symmetric, ad-invariant bilinear form on the algebra $\Lie(G)$.}
\be
\L = -\ha \h \, \Tr[ J_+ J_-] \ , \qquad J_\pm = g^{-1} \del_\pm g \ , \quad g\in G\ . \la{PCML}
\ee
In that case the current appearing in the Lax connection is just the Maurer-Cartan current, $\mc A_\pm = J_\pm$. More generally, e.g.\ for integrable deformations of the PCM, the currents  take a model-dependent form $\mc A_\pm=\mc A_\pm(g)$  in terms of the physical fields, schematically denoted $g$.
 
The curvature of the Lax \rf{fc} is
\be
\begin{aligned}
F_{+-}(L) &:= \del_+ L_- - \del_- L_+ + [L_+, L_-] \\
&= \frac{1}{1-z^2} \, \del \mc A + \frac{z}{1-z^2} \, F_{+-}(\mc A) \ .
\end{aligned}
\ee
The equations of motion are assumed to be encoded in the vanishing of the curvature for all $z$, i.e.\ the flatness and conservation of $\mc A$,
\be
\del \mc A = 0 \ , \qquad F_{+-}(\mc A)=0 \ . \la{os}
\ee
Hence all theories with this same Lax connection 
 have this same  set of `on-shell' equations.  
 
Some combinations  of these equations may be Bianchi identities, which hold even off-shell. For example for the PCM we have $\mc A_\pm \equiv g^{-1} \del_\pm g$ implying $F_{+-}(\mc A) \equiv 0$. In general the currents $\mc A_\pm(g)$ may solve one or more Bianchi identities (which we denote with the symbol $\equiv$)  of the form\foot{See \cite{HL} for discussion of Bianchi identities of this form in theories related to the PCM.}
\be
P(g) \cdot F_{+-}(\mc A)  \equiv O(g) \cdot (\del \mc A)  \ , \la{Bia}
\ee
where the linear maps $O(g),P(g):\Lie(G) \to \Lie(G)$ may depend on $g$. 
 
We propose that, for all  theories with Lax connections of the from \rf{fc} satisfying the Bianchi Completeness Assumption \ref{BCA} below,  the 1-loop divergences take a universal form in terms of $\mc A$.
Indeed the 1-loop renormalization is fully determined by the classical equation of motion, and these theories have the same equations of motion, up to
the distinction between Bianchi identities and equations of motion. As we shall argue below, that distinction does not affect the 1-loop on-shell divergences.

Hence, at the 1-loop level, one may treat any model of this type like the PCM: treating $F_{+-}(\mc A) \equiv 0$ as a Bianchi identity and $ \del \mc A = 0$ as an equation of motion.  Then the  divergences are  simply the same as the PCM,  with the Maurer-Cartan current $J_\pm$ replaced with the appropriate Lax current $\mc A_\pm$. The resulting prescription for the 1-loop effective Lagrangian is 
\be
\ddt \widehat\L^{(1)} = \ddt \widehat\L_{\rm PCM}^{(1)}\Big|_{J\to \mc A} =-\ha \cG \Tr[ \mc A_+ \mc A_- ] \ . \la{1p}
\ee
Here we have used the fact the only 1-loop  divergences of the PCM \rf{PCML} are absorbed into its coupling as $\ddt \h = \cG$, so its 1-loop effective Lagrangian is $\ddt \widehat\L_{\rm PCM}^{(1)} =-\ha \cG \Tr[ J_+ J_- ]$.
In the following sections we will explicitly derive the result \rf{1p} by a path integral argument.

\subsection{General argument: Bianchis vs.\ EOMs} \la{gen}
In this section we give a path integral argument that, under certain assumptions, 1-loop on-shell divergences do not depend on which equations are Bianchi identities and which are equations of motion. We will consider a general 2d field theory with Lax connection $L_\pm = L_\pm(z; \mc A(g))$ in terms of some currents $\mc A_\pm(g)\in \Lie(G)$ that depend on the physical fields $g$.
We expand the currents around a  background,
\be
\mc A_\pm = \bar{\mc A}_\pm + a_\pm \ , \quad  \qquad \bar{\mc A}_\pm = \mc  A_\pm(\bar g) \ , \la{ax}
\ee
where $\bar g$ is an on-shell background for $g$. We assume the `completeness' of the Bianchi identities:
 \vspace{-0.4cm}
\noindent{\begin{center} \fbox{\begin{minipage}{0.99\textwidth}\begin{BCA}\label{BCA} \ \\
Assume that, to leading order in the background field expansion, a subset of the zero-curvature equations is equivalent to the Lax currents taking the correct form $\mc A_\pm=\mc A_\pm(g)$ in terms of the physical fields.
\end{BCA}\end{minipage}}\end{center}}
 \vspace{0.2cm}
\noindent In other words, some subset\foot{In particular, we are assuming that the zero-curvature equations $F_{+-}(L)=0$ include  \textit{all} necessary Bianchi identities. Note that this assumption seems to single out $\s$-models, and excludes models like sine-Gordon, which would require extra Bianchi identities to be imposed by hand. See appendix \ref{A} for sufficient conditions for Bianchi completeness.}  of the Lax zero-curvature equations are the Bianchi identities,\foot{One might worry about the possibility of a more general (e.g.\ non-linear) relation between the Bianchi identities and the zero-curvature equations than \rf{Bfo}, but in any case the linearized Bianchi identities would still take the form \rf{new}.\la{fos}}
\be
\text{Bianchi}_\a (g,A) \equiv B_{\a}{}^I(g)  \ \text{zero-curv}_{I}(\mc A)\equiv 0 \ , \la{Bfo}
\ee
definded by the property that the physical currents satisfy them off-shell, $\text{Bianchi}_\a (g,\mc A(g)) \equiv 0$. Here $\text{zero-curv}_{I}(\mc A)= 0$ are all of the equations (indexed by $I$) following from the zero-curvature condition $F_{+-}(L)=0$, and $B_{\a}{}^I(g)$ is some matrix depending on $g$. 

The linearized Bianchi identities are then
\be
\text{Bianchi}_\a (\bar g) \cdot a \equiv   B_{\a}{}^I(\bar g)  \ \text{zero-curv}_{I} \big(\mc A(\bar g) + a \big)\Big|_{\O(a)}  \equiv 0 \ , \la{new}
\ee 
where the symbol  $|_{\O(a)}$ indicates truncation at linear order in $a$. We may indeed replace \mbox{$B_{\a}{}^I(g) \to B_{\a}{}^I(\bar g)$} to linear order in \rf{new} since the background field is on-shell, $\text{zero-curv}_I\big(\mc A(\bar g) \big)=0$.\sloppy

The Bianchi Completeness Assumption is that the linearized Bianchi identity \rf{new} is equivalent to the fluctuations $a_\pm$ of $\mc A_\pm$ arising from fluctuations $\phi$ of the physical fields $g$,
 \be
\mc A_\pm(\bar g) + a_\pm = \mc A_\pm(\bar g + \phi) \ . \la{phys}
 \ee

\bigskip

The 1-loop effective action is given by
\be
\widehat S^{(1)}(\bar g) = -i \log \int \mathcal D \phi^i \  e^{i \int \phi^i \, \O_{ij}(\bar g)\,  \phi^j} = \tfrac{i}{2} \log \det\, {\O(\bar g)} \ ,\la{eff1}
\ee
where $\phi^i \, \O_{ij}(\bar g)\,  \phi^j =\L(\bar g + \phi)-\L(\bar g) + \O(\phi^3)$ is the linearized Lagrangian to quadratic order, and the linearized equations of motion are $\text{EOM}_i(\bar g) \cdot \phi \equiv \O_{ij}(\bar g) \, \phi^j=0$.

We observe that \rf{eff1} has an alternate path integral representation,\foot{We denote the operators characterizing the linearized equations of motion, Bianchi identities and zero-curvature equations by $\text{EOM}_i$, $\text{Bianchi}_\a$ and  $\text{zero-curv}_I$, where $i,\a$ are some indices and $I=(i,\a)$.}
\be
2\widehat S^{(1)}(\bar g) = - i \log \int \mathcal D u^i  \, \mathcal D \phi^i \  e^{i \int u^i \, \O_{ij}(\bar g)\,  \phi^j}    = - i \log \int \mathcal D u^i  \, \mathcal D \phi^i \  e^{i \int u^i \, \text{EOM}_i(\bar g) \cdot \phi } \la{L1}
\ee
The extra factor of 2 in the left-hand-side of \rf{L1} arises from the identity,
\be
\log \det \begin{pmatrix}
0 & \O(\bar g) \\
\O(\bar g)^T & 0  
\end{pmatrix}
= 2 \log \det \, {\O(\bar g)} + \text{constant} \ ,
\ee
and we drop the finite constant contribution.
One may think of eq.\ \rf{L1} as representing the linearized equation of motion as a Bianchi identity imposed by the Lagrange multipliers $u^i$.

Now let us `integrate in' the Bianchi identities to make manifest the symmetry between equations of motion and Bianchi identities. Due to the Bianchi Completeness Assumption \ref{BCA}, we may trade the integral over $\phi$ for an integral over the fluctuations $a_\pm$ of the Lax currents 
while imposing the linearized Bianchi identities through delta-functions,
\unskip\foot{We schematically denote the linearized equations of motion both in the $\p$ variable and the $a$ variable as $\text{EOM}_i$; indeed the equations of motion can  be written in terms of the fluctuation $a$ of the Lax current since they follow from a zero-curvature equation.  We may neglect the determinant from the change of variables $ a \leftrightarrow (\phi^i, \text{Bianchi}_\a(\bar g)\cdot a)$ since it is of first-order in derivatives, so the resulting counterterms are finite and will only contribute to 2-loop and higher divergences. }
\begin{align} \la{fi0}
2\widehat S^{(1)}(\bar g) &= - i \log \int \mathcal D u^i  \, \mathcal D a_\pm \ \delta(\text{Bianchi}_\a (\bar g)  \cdot a ) \  e^{i \int u^i \, \text{EOM}_i(\bar g) \cdot a} \\
&= - i \log \int \mathcal D u^i \, \mathcal D v^\a \, \mathcal D a_\pm \   \exp{i \int (u^i \, \text{EOM}_i(\bar g)  + v^\a \, \text{Bianchi}_\a (\bar g) ) \cdot a } \ .\la{fi}
\end{align}
In \rf{fi} we have recast the delta-function constraints using Lagrange multipliers $v^\a$.

Eq. \rf{fi} is clearly invariant under ($\bar g$-dependent) rotations between the equations of motion and Bianchi identities, since these may be absorbed into a change of variables of $u$ and $v$. By assumption the equations of motion and Bianchi identities are related to the zero-curvature equations by such a rotation, so we obtain the result
\be
2 \widehat S^{(1)}(\bar{\mc A}) = - i \log \int \mathcal D U^I \, \mathcal D a_\pm \   \exp{i \int U^I \ \text{zero-curv}_I (\bar{\mc A}) \cdot a }  \ . \la{res}
\ee
Note that \rf{res} only depends on $\bar{\mc A}$ (not explicitly on $\bar g$) since the Lax connection and its zero-curvature equations only depend on $\mc A$.

 Thus we conclude that, for a given analytic structure of the Lax connection, all theories satisfying the Bianchi Completeness Assumption \ref{BCA} have the same 1-loop on-shell effective action $\widehat S^{(1)}(\mc A)$ in terms of the Lax currents $\mc A$. 
This argument is valid assuming the absence of gauge redundancies (including, e.g., $\ka$-symmetry of the Green-Schwarz string), which would need to be appropriately fixed.

\subsection{Divergences for PCM and related models}

 In the particular case of
  Lax connections of the PCM-form \rf{fc}, we write  $U=(u,v)\in\Lie(G) \oplus \Lie(G)$ so that \rf{res} becomes
\be
2 \widehat S^{(1)}(\bar{\mc A}) = - i  \log \int \mathcal Du \, \mathcal D v\, \mathcal D a_\pm \   \exp{i\int  \Tr \big[ u \, \del (\bar{\mc A} + a)  +  v\,  F_{+-}(\bar{\mc A}+a) \big]  \Big|_{\O(a)}}\ . \la{zs}
\ee
Thus the only theory-dependence comes from the form of the Lax current $\mc A_\pm(g)$ in terms of the physical fields. This leads to the prescription \rf{1p}: the 1-loop effective action for any theory of this type is obtained from  that of the PCM by replacing $J_\pm \to \mc A_\pm(g)$.

\bigskip
We can also compute the 1-loop effective action directly from \rf{zs} using Feynman diagrams. The following computation will therefore manifestly treat the Bianchi identities and equations of motion on an equal footing, and treats the Lax currents $\mc A_\pm$ (or their fluctuations $a_\pm$) as the fundamental fields, as well as the Lagrange multiplier fields $u,v$.

\begin{figure}[H]
\centering
\raisebox{-0.5\height}{\includegraphics[scale=0.11]{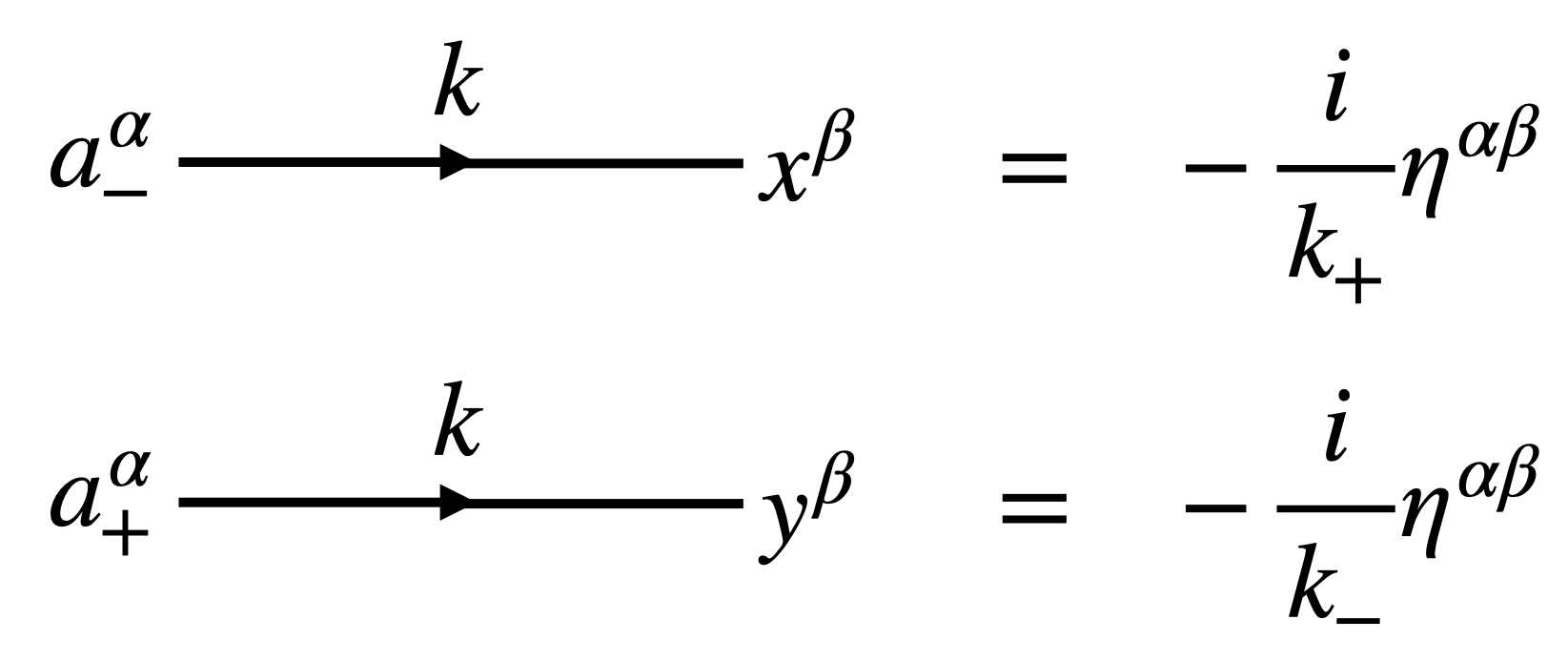}} 
\raisebox{-0.5\height}{\includegraphics[scale=0.11]{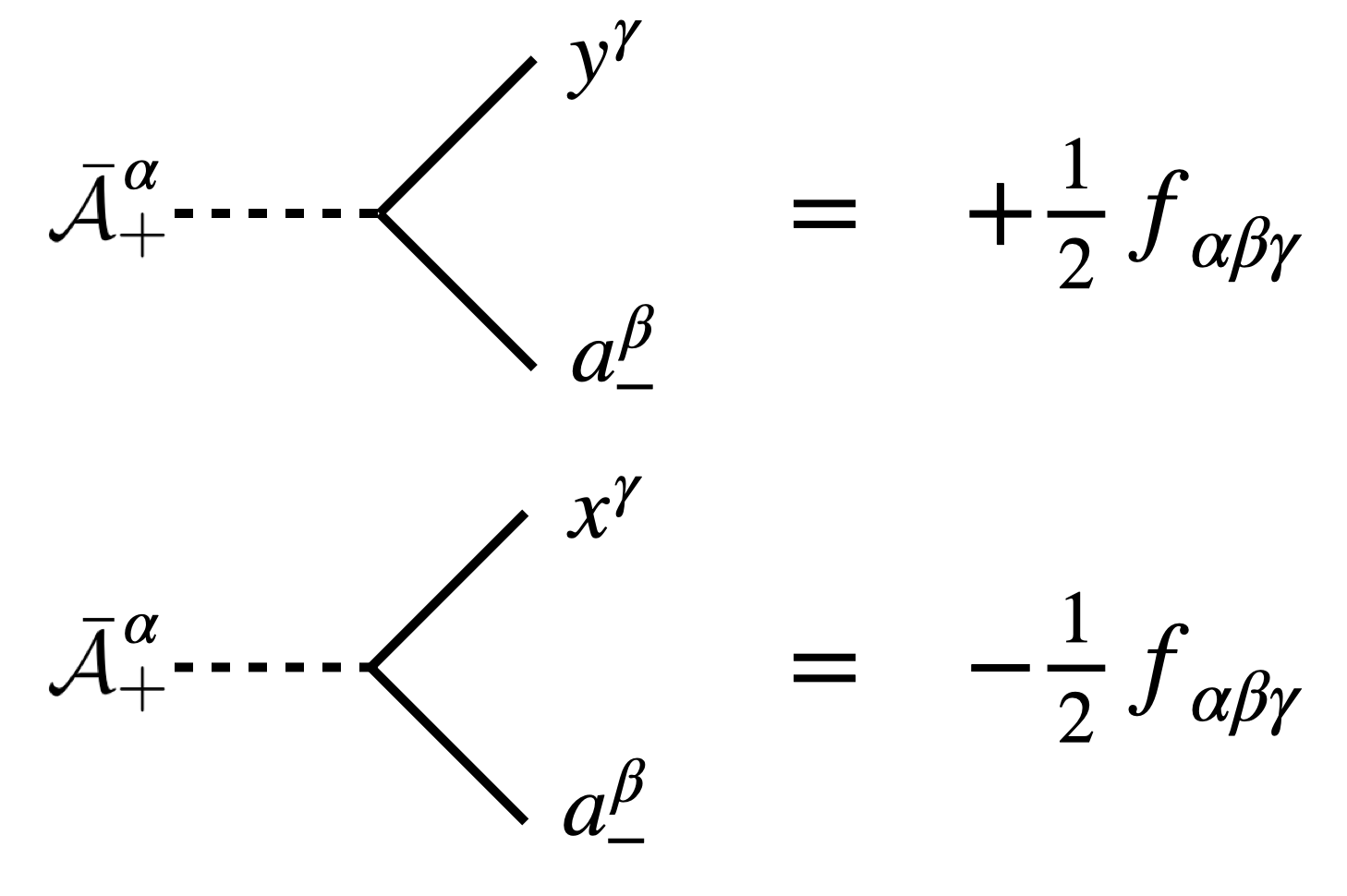}\hspace{1.7cm}\includegraphics[scale=0.11]{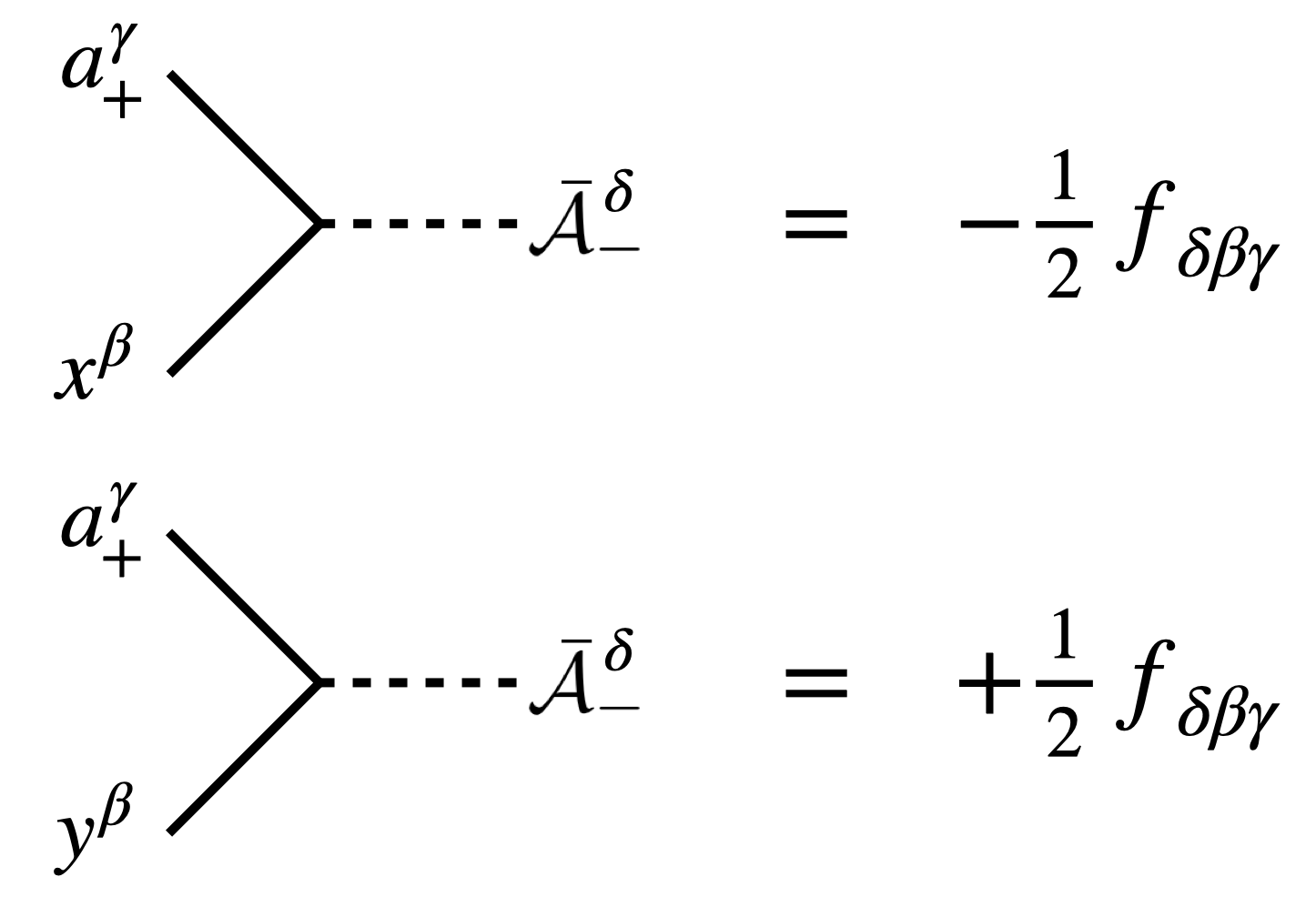}} 
\caption[] {\small The Feynman rules following from the path integral \rf{vec}. Fluctuation fields and background fields are denoted by solid and dashed lines respectively. $k_\pm$ denote the light-cone components of the 2d momenta. The indices $\a,\b,\g$ label the Lie algebra generators $T_\a$, the structure constants are defined by $[T_\b, T_\g] = i f^\a{}_{\b\g} T_\a$ and indices are raised and lowered by $\eta_{\a\b} = \Tr[ T_\a T_\b]$ and its inverse $\eta^{\a\b}$ (see conventions in appendix \ref{Apeq}).\label{Pfeyn}}
\end{figure}

Using the fact that the background is on-shell ($\del \bar{\mc A}=F_{+-}(\bar{\mc A})=0$), truncating at quadratic order in the fluctuations $a_\pm,u,v$ and changing variables $(u,v) \to (x,y)=(u+v,u-v)$ we get from \rf{zs}
\be \begin{aligned}
2 \widehat S^{(1)} = &- i  \log \int \mathcal Dx \, \mathcal D y\, \mathcal D a_\pm \ \\
  &\qquad \qquad \exp{}i\int  \Tr \Big[  x \, \del_+a_-  +  y \, \del_- a_+ + \ha (x-y)\big( [\bar{\mc A}_+, a_-] +[ a_+,\bar{\mc A}_- ]\big)\Big]  \ . \la{vec}
\end{aligned} \ee
Now one can  quantize this system of scalars $x$, $y$ and 2d vectors $a_\pm$.\foot{Equivalently one could  change variables $a_+ = \del_+ w$, $a_- = \del_- z$ to exchange the vector fluctuations for scalars --- the result is exactly the same. Note that the determinant for such a 1st order change of variables is finite so would not contribute to 1-loop divergences.} 
The  Feynman rules resulting from eq.\ \rf{vec} are indicated in Figure \ref{Pfeyn}.

\begin{figure}[H]
\centering
\raisebox{-0.5\height}{\includegraphics[scale=0.14]{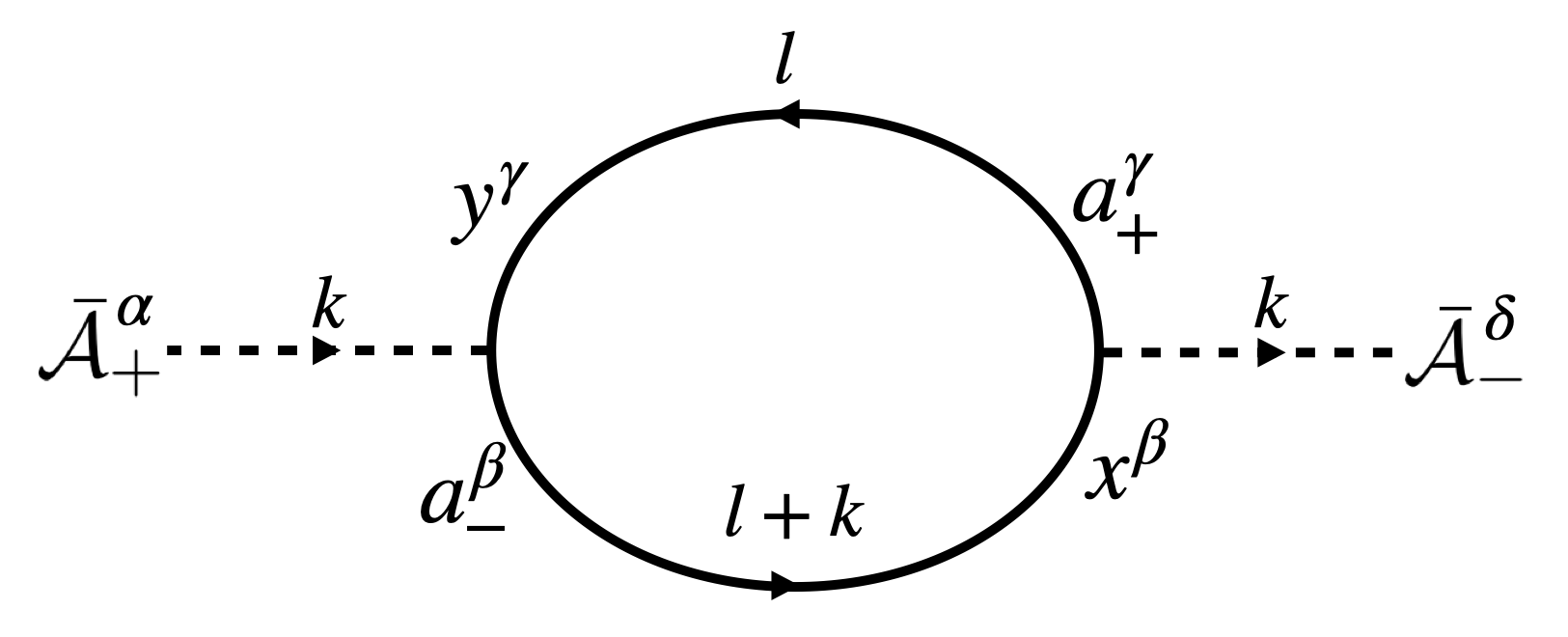}}
\caption[] {\small The only divergent diagram contributing to the 1-loop effective action \rf{1lP} for theories with Lax connection of the form \rf{fc}. \label{Pdiag}}
\end{figure}

 There is only one divergent diagram, indicated in Figure \ref{Pdiag}, which contributes
 \be
2 \widehat S^{(1)} = \Big(\int \frac{d^2 l}{(2\pi)^2} \frac{1}{l_-(l+k)_+} \Big)  \ \tfrac{1}{4} f_{\a\b\g}f_\d{}^{\b\g} \int  d^2 \xi  \ \bar{ \mc A}^\a_+ \bar{ \mc A}^\d_- 
 \ee
The  integral over $l$  diverges as $-\tfrac{1}{2\pi} \log\tfrac{\Lambda}{\m}$  where $\Lambda$ is a UV momentum cutoff and $\mu$ is the RG scale. Using the identity $f_{\a\b\g}f_\d{}^{\b\g} = -2\cG \eta_{\a\b}$  we confirm the result  \rf{1p},
\be \begin{aligned}
&\widehat S^{(1)} = \tfrac{1}{4\pi} \int d^2 \xi \ \widehat \L^{(1)}  = \tfrac{1}{4\pi} \int d^2 \xi \  \log\tfrac{\Lambda}{\m} \, \ha \cG \Tr[ \bar{ \mc A}_+ \bar{ \mc A}_- ] + \text{finite} \\
&\ddt \widehat\L^{(1)} = -\ha \cG \Tr[ \bar{ \mc A}_+ \bar{ \mc A}_- ] \qquad (t\equiv \log \m) \ . 
\end{aligned} \la{1lP}\ee

\subsection{Examples}
We will now show that the result \rf{1p} matches the known $\b$-functions in various examples with Lax connections of the form \rf{fc}.
As we prove in appendix \ref{A}, the Bianchi Completeness Assumption \ref{BCA} is satisfied  and hence the argument above valid for all of the following examples. 
More generally, it is satisfied whenever the Bianchi identities take the form $F_{+-}(g^{-1} \del g)\equiv 0$, with $g$ a group-valued physical field and the Lax currents related by $A_\pm(g) = O_\pm(g) \, g^{-1} \del_\pm g$ where the operators $O_\pm(g)$ are invertible. For algebra-valued  fields  $v$,  the leading order form of that condition with $g\sim 1+v$ ($v$ small) is also sufficient for Bianchi completeness.

\paragraph{PCM with WZ term.} 
For the PCM with a WZ term with arbitrary level $k$ (integer for compact groups),
\be
\L = -\ha { \rm h}  \, \Tr[ J_+ J_-] + k \, \L_{\rm WZ} (g) \ , \qquad \quad J=g^{-1} \del g \ , \qquad g\in G \ , \la{PCMk}
\ee
the flat, conserved Lax current in \rf{fc} is $\mc A_\pm = (1\pm \tfrac{k}{h}) J_\pm $. The  result \rf{1p} then gives
\be
\frac{d}{dt} \widehat\L^{(1)}  = - \ha \cG (1-(\tfrac{k}{\rm h})^2) \, \Tr[ J_+ J_-] \ , 
\ee
which is equivalent to the standard 1-loop $\b$-functions for the couplings $\rm h$, $k$ in \rf{PCMk},
\be
\ddt {\rm h }= \cG (1-(\tfrac{k}{\rm h})^2) \  , \qquad \quad \ddt k = 0 \ .
\ee

\paragraph{$\eta$-deformation.} The $\eta$-deformation of the PCM is \cite{klim}
\be
\L = -\ha {\rm h} \, \Tr[ J_+ \frac{1}{1-\eta \mc R_g} J_- ] \ ,  \la{etam}
\ee
where $\mc R$ is a linear map on the algebra, antisymmetric with respect to $\Tr$, and solving the modified classical Yang-Baxter equation,
\be
[\mc R u  , \mc R v]  - \mc R([\mc R u  , v] +[\mc R u  ,  v] ) + c^2 [u,v] = 0 \ , \qquad u,v\in \Lie(G) \ , \la{YBE}
\ee
and we denote $\mc R_g := \Ad_{g^{-1}} \mc R \Ad_g$.  The parameter $c$ in \rf{YBE} may be fixed to $0$, $1$ or $i$ by real rescalings of the $\mc R$.

The flat, conserved Lax current is $\mc A_\pm = (1-c^2 \eta^2) \frac{1}{1\pm \eta {\mc R_g}} J_\pm $. The result \rf{1p} then gives
\be
\frac{d}{dt} \widehat\L^{(1)} = - \ha \cG (1-c^2\eta^2)^2 \, \Tr[ J_+ \frac{1}{(1 - \eta {\mc R_g})^2} J_- ] \ .
\ee
One may check that this counterterm is indeed equivalent to the standard running of the couplings $\rm h$, $\eta$ in \rf{etam} \cite{Squellari,SST}
\be
\ddt \begin{pmatrix}
{\rm h}\\ \eta
\end{pmatrix} = \begin{pmatrix}
{\rm h}\\ \eta
\end{pmatrix}
\frac{\cG}{\rm h} (1-c^2\eta^2)^2 \ .
\ee

\paragraph{$\l$-deformation.} The $\l$-deformation on the group $G$ is \cite{lambda}
\be \begin{aligned}
&\L = k \Big(\L_{\rm WZW}(g) + \Tr[ J_+ A_- - K_- A_+ + g^{-1} A_+ g A_- - \l^{-1} A_+ A_- ] \Big) \ , \\
&\qquad \qquad  J_+ = g^{-1} \del_+ g \ , \quad K_- = \del_- g g^{-1} \ .
\end{aligned} \ee
Integrating out the auxiliary 2d gauge field $A_\pm$,
 \begin{align}
 A_+(g) = -(\Ad_g^{-1}-\l^{-1})^{-1} J_+ \ , \qquad  A_-(g) = (\Ad_g-\l^{-1})^{-1} K_- \ . \la{osv}
 \end{align}
the theory may be written as a conventional  $\s$-model, 
\be
 \L 
 =k \Big(\L_{\rm WZW}(g) + \Tr[ J_+ \frac{1}{\Ad_g - \l^{-1}} K_- ] \Big) \ .
 \ee
The flat, conserved Lax current is $\mc A_\pm = \tfrac{2}{1+\l} A_\pm(g)$, with $A_\pm(g)$ denoting the on-shell value  \rf{osv} of the gauge field. The result \rf{1p} then gives
\be
\frac{d}{dt} \widehat\L^{(1)} = - \frac{2 \cG}{(1+\l)^2} \Tr[ J_+ \frac{1}{(\Ad_g - \l^{-1})^2} K_- ]  \ , 
\ee
which may be absorbed into the standard running of the coupling $\l$ \cite{Itsios:2014lca,CLa,ah},
\be
\ddt k = 0 \ , \qquad \ddt \l = - \frac{2 \cG}{k}\Big(\frac{\l}{1+\l}\Big)^2 \ .
\ee

\paragraph{Non-abelian T-dual of PCM.} Let us consider the non-abelian T-dual of the PCM. First we consider the interpolating model (obtained by gauging the left symmetry $G_L$, imposing with Lagrange multiplier $v$ that the gauge field is pure gauge, and gauge fixing $g=1$),
\be
\L_{\rm int} = -\ha \,  \Tr[{ \rm h} \, A_+A_- + v  \, F_{+-}(A) ]  \ . 
\ee
On one hand integrating $v$ gives back the PCM with $A=g^{-1}\del g$ identified as the flat, conserved current $\mc A$. On the other hand integrating the  gauge field to get $A_\pm =  \frac{1}{ \ad_v \pm  {\rm h}} \del_\pm v$, we find the non-abelian T-dual theory (here $v\in \Lie(G)$ is a scalar field),
\be
\L_{\rm NATD} =  \ha \,  \Tr[ \del_+ v \frac{1}{\ad_v - \rm h} \del_- v ] \ . \la{NATD}
\ee
This path integral transformation is in particular a canonical transformation, so the NATD inherits the Lax connection \rf{fc} of the PCM, with the flat, conserved Lax current identified with the on-shell value of $A$,
\be
\mc A_\pm = \frac{1}{ \ad_v \pm  {\rm h}} \del_\pm v \ .
\ee
Thus the result \rf{1p} predicts the 1-loop RG flow to be
\be
\ddt \widehat\L^{(1)}_{\rm NATD} = \ha \cG\, \Tr[ \del_+ v \frac{1}{(\ad_v - \rm h)^2} \del_- v ] \ ,
\ee
which is equivalent to just $\rm h$ running in \rf{NATD} with the standard PCM $\b$-function, $\dt \h = \cG$.

\paragraph{Pseudo-dual of PCM.} The PCM admits a classical `pseudo-dual', by exchanging the roles of the equation of motion and Bianchi identity in \rf{os}. Rather than  $F(\mc A)\equiv 0$, let us instead take $\del \mc A \equiv 0$ to be the Bianchi identity. We may explicitly solve it as
\be
{\mc A}_a = \l \, \e_{ab} \,  \del^b \phi \ , \la{ps}
\ee
where we have inserted the coupling $\rm \l$ for convenience. The other equation $F(\mc A)=0$, which we will treat as an equation of motion, then becomes $ \del^2 \p - \l \, \e^{ab}[\del_a \p, \del_b \p ]  = 0$. Remarkably it follows from a dual Lagrangian \cite{ZM},\foot{The same model \rf{pseudo} had previously been considered in \cite{Lund}.}
\be
\L_{\rm pseudo} = - \, \Tr\big[ \tfrac{1}{\l} (\del\p)^2 +  \tfrac{2}{3} \e^{ab} \p [\del_a \p, \del_b \p ] \big] \ . \la{pseudo}
\ee
The PCM and its psuedo-dual are equivalent at the level of equations of motion: there is a  map between classical solutions and they share the same Lax connection \rf{fc}. The flat, conserved current of the pseudo-dual is  given by \rf{ps}, so the result \rf{1p} gives the following 1-loop divergences
\be
\ddt  \widehat\L^{(1)}_{\rm pseudo} = -\ha  \cG \, \Tr[(\l \, \e \, \del \p)^2] = \ha \cG \l^2 \, \Tr[ (\del \p)^2] \ .
\ee
Indeed this implies the correct result for the pseudo-dual's 1-loop $\b$-function \cite{Nappi},
\be
\ddt \l = \ha \cG  \l^4 \ .
\ee
This $\b$-function has the opposite sign to the PCM, since the free limit $\l=0$ is obtained in the IR instead of the UV. This was seen as evidence for the quantum inequivalence of the two models \cite{Nappi}, but it is interesting that, in terms of their Lax currents, their 1-loop $\b$-functions coincide and are given by the formula \rf{1p}.
In this sense, their classical equivalence may be seen as extending to the level of 1-loop divergences.\foot{One can also see directly that the linearized equations of motion (and thus the 1-loop divergences) of the PCM and its pseudo-dual are the same. A standard quantization involves linearizing the equations of motion under fluctuations of the physical fields (uniformly satisfying the Bianchi identities). For the PCM, $\delta g = g Y$ gives $\delta \mc A_\pm = [\bar{\mc A}_\pm, Y] + \del_\pm Y$ and the equaton of motion $\del \mc A =0$ has the linearized form $\del^2 Y + [\bar{\mc A}, \del Y] =0$ (contracting 2d indices with $\eta^{ab}$ inside the commutator). For the pseudo-dual, $\delta \phi = \l^{-1} Y$ gives $\delta \mc A_\pm = \pm \del_\pm Y$  and the equation of motion $F(\mc A)=0$  has the same linearized form $\del^2 Y + [\mc A, \del Y]=0$. One also obtains the same linearized equation from the PCM with WZ term.}

\section{\texorpdfstring{`Pure-spinor' $\mathbb{Z}_T$}{'Pure-spinor' ZT} cosets and deformations \la{CD}}
The path integral argument in section \ref{gen} applies in principle to any analytic structure of the Lax connection (i.e.\ any choice of poles in the spectral $z$ plane). For each analytic structure, there should correspond a universal formula for the 1-loop on-shell divergences $\ddt \widehat{\L}^{(1)}(\mc A)$ in terms of the currents $\mc A$ (which are the residues of the poles). 

In this section we focus on the analytic structure
 \be
 L_\pm = \sum_{j=0}^{T-1} z^{\pm j} \, P_{\mp j} \mc A_\pm \ , \la{ZTL}
 \ee 
where the current $\mc A_\pm$ is valued in a Lie (super)algebra $\Lie(G)$ assumed to admit a $\mathbb{Z}_T$ grading, and $P_{j\ (\text{mod } T)}$ are the corresponding projectors onto the graded subspaces $\mathfrak{g}_j$ of  $\Lie(G)$.\foot{For superalgebras we assume the even/odd graded generators to be bosonic/fermionic respectively, implying that $T$ is even. See appendix \ref{Apeq} for further conventions.}
 
The propotypical theories with Lax connections of this form are the integrable $\mathbb{Z}_T$ cosets of `pure-spinor' type  \cite{Berkovits,Young} (with Lax current $\mc A_\pm = J_\pm$),
\be
\L = -\ha \h \, \Tr[ J_+ \Big( \sum_{j=0}^{T-1} j P_j \Big) J_- ] \ .  \la{ZT}
\ee
Here $\Tr$ is a symmetric, ad-invariant bilinear form (for superalgebras we may take it to be the supertrace).
The name `pure-spinor' derives  from the $T=4$ case, where such models on supergroups (consistently coupled to worldsheet ghosts) describe pure-spinor string theory on supercosets. We emphasize that the following discussion does not apply to Green-Schwarz type models \cite{GS,Hoare}, which have a different Lagrangian to \rf{ZT} (see discussion in section \ref{out} below).\foot{See also \cite{osten} for more general classes of integrable coset models.}

\bigskip

There are two subtleties to be understood  before applying the argument of section \ref{gen} to Lax connections of this form. First, the currents $P_{\mp i} \mc A_\pm$ multiplying poles of different orders in \rf{ZTL}  are constrained to be in subspaces of the algebra. This is not problematic, as we can simply understand this constraint to be imposed in the integrals $\mc{D}a_\pm$ at every stage of the argument in section \ref{gen}.

Second,  equations of motion corresponding to \rf{ZTL}  are invariant under an apparent gauge transformation under the 
 zero-graded subgroup $G_0$,
\be
\mc A_\pm  \to  h^{-1} \mc A_\pm h + h^{-1} \del_\pm h
\ , \qquad \quad h \in G_0\ , \la{Lg} 
\ee 
and so do not depend on all of the components of $\mc A$. We restrict to theories that do indeed have a local $G_0$ gauge invariance with the Lax current transforming in this way. The argument of section \ref{gen} is then easily adapted by  gauge fixing  on the fluctuation fields $\phi$ in eq.\ \rf{L1}. One may choose a gauge that corresponds to $P_0 \, a_+ =0$ upon changing variables to the fluctuations $a_\pm$ of $\mc A_\pm$ in \rf{fi}. In this axial-type gauge the ghosts decouple and the 1-loop effective action is then given by (cf.\ \rf{res}),
\be
2 \widehat S^{(1)}(\bar{\mc A}) = - i \log \int \mathcal D U^I \, \mathcal D a_\pm  \ \d(P_0 \, a_+)\   \exp{i \int U^I \ \text{zero-curv}_I (\bar{\mc A}) \cdot a }  \la{resT}
\ee

Denoting $\mc A^j \equiv P_j \mc A$, the zero-curvature equations following from the Lax connection \rf{ZTL} are
\begin{align}
& \del_+ \mc A_-^k+ \sum_{k\leq j<T}[\mc A_+^{T+k-j},\mc A_-^{j}]  = 0  \ ,  \qquad \del_- \mc A_+^k-  \sum_{0\leq \hat \jmath<k}[\mc A_+^{k-\hat \jmath},\mc A_-^{\hat \jmath}]  = 0  \ ,  \\
&F_{+-}(\mc A^0)+\sum_{0<j<T} [\mc A_+^{T-j},\mc A_-^{j}]  =  0 \ \ \ \    \qquad (0<j,k<T \ , \ \  0\leq \hat \jmath, \hat k<T) \ .
\end{align}
Linearizing these equations around a background ($\mc A= \bar{\mc A} +a$), fixing the gauge $a^0_+ = 0$ and substituting into \rf{resT}, we obtain 
\begin{align}
&2 \widehat S^{(1)}= - i \log \int \mathcal D x^{\hat{k}} \, \mc D y^k \, \mathcal D a_+^k \, \mathcal D a_-^{\hat k} \no\\
&\qquad \qquad \qquad \exp{i \int \Tr\Big[  x^{T-\hat k}\Big(\del_+  a_-^{\hat k}+ \sum_{\hat k\leq j} [\bar{\mc A}_+^{T+\hat k-j}, a_-^{j}] + \sum_{\hat k<j}[ a_+^{T+\hat k-j},\bar{\mc A}_-^{j}] \Big)}  \la{ZTpa}\\
&\qquad \qquad  \qquad \qquad\quad  \qquad +  y^{T- k}\Big(\del_-  a_+^{ k} - \sum_{ \hat \jmath<k}\big( [\bar{\mc A}_+^{ k-\hat \jmath}, a_-^{\hat \jmath}] + [ a_+^{k-\hat \jmath},\bar{\mc A}_-^{\hat \jmath}] \big) \Big)\Big]  \ . 
\no
\end{align}
Here we are summing over repeated indices $0<j,k<T$ and $0\leq\hat{\jmath}, \hat{k}<T$ and we have written $U^I=(x^{\hat k},y^k)$ with $x^{\hat k}\in \mathfrak{g}_{\hat k}$, $y^{ k}\in\mathfrak{g}_{ k}$.

\begin{figure}[H]
\vspace{1cm}
\centering
\raisebox{-0.5\height}{\includegraphics[scale=0.12]{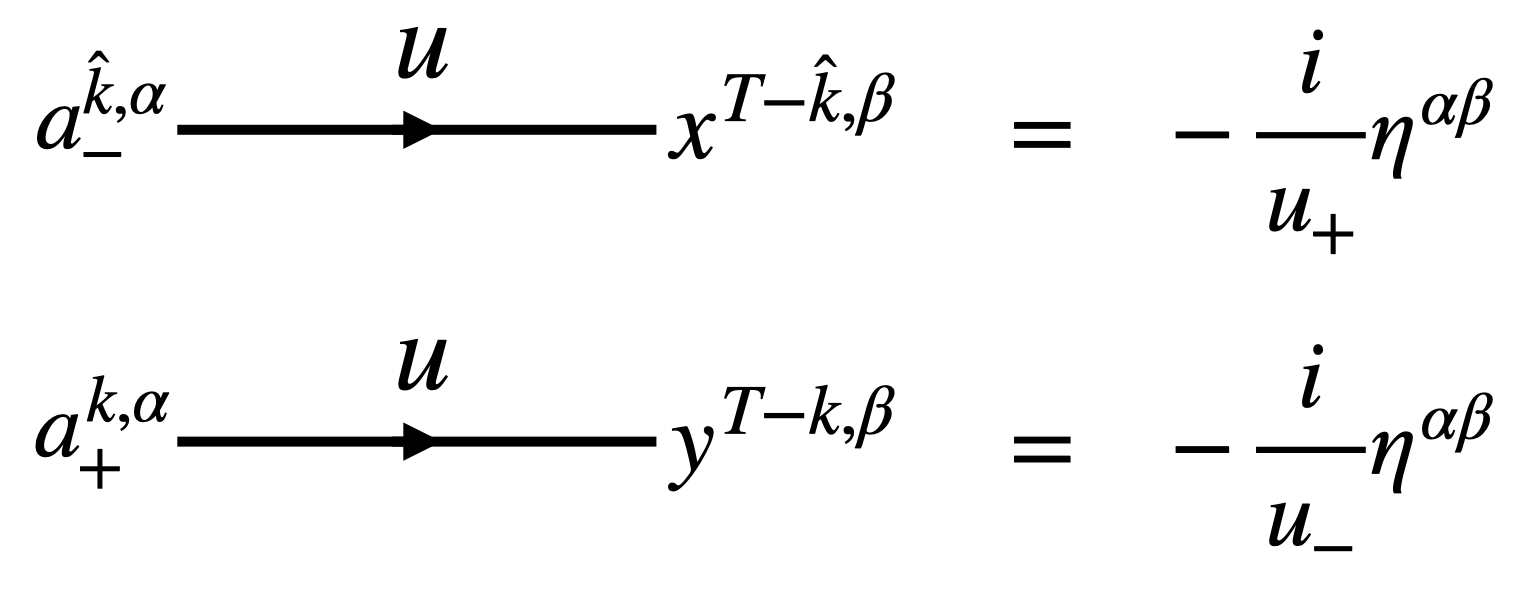}} 
\raisebox{-0.5\height}{\includegraphics[scale=0.12]{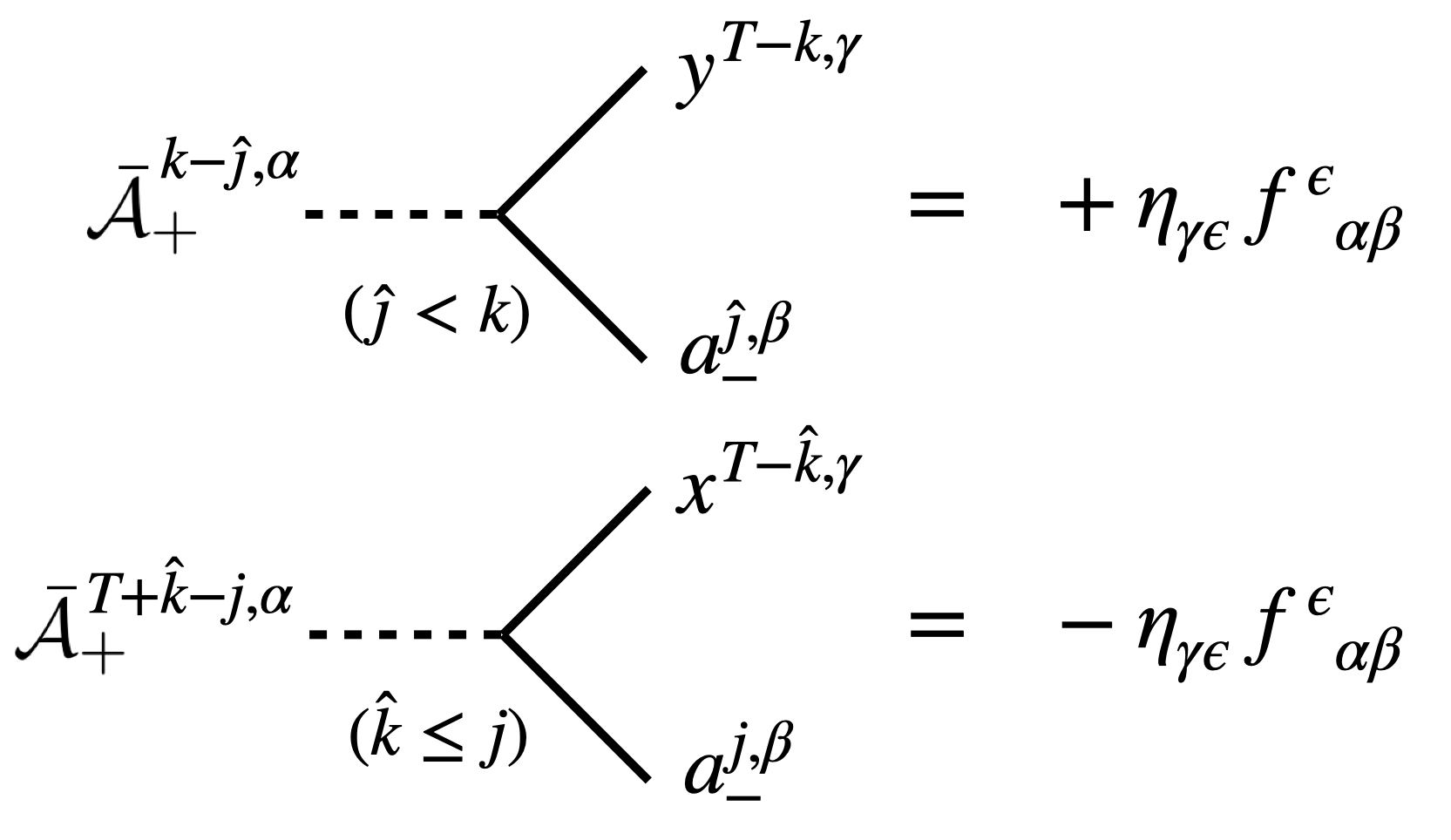}\hspace{1.7cm}\includegraphics[scale=0.12]{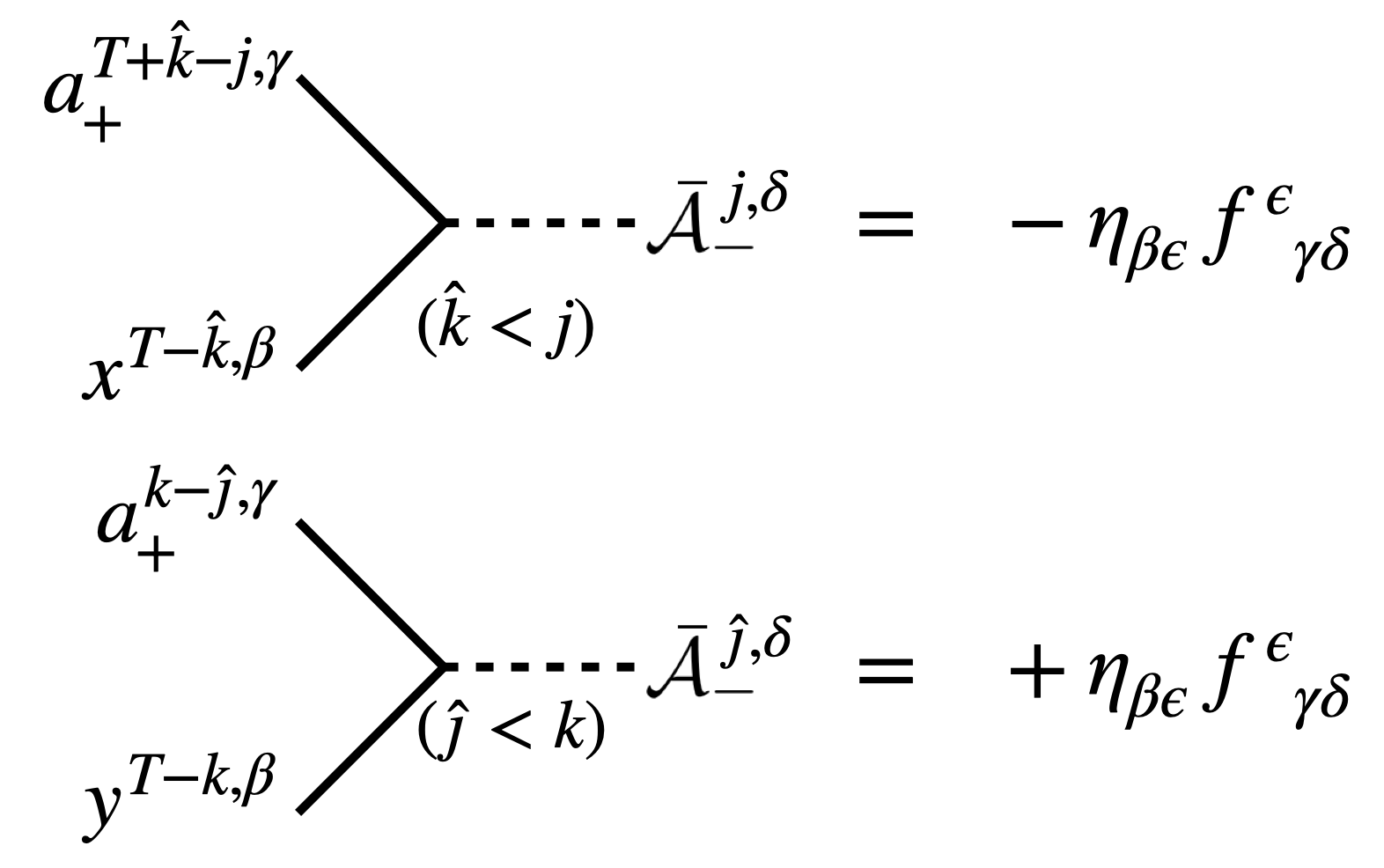}} 
\caption[] {\small The Feynman rules following from the path integral \rf{ZTpa}. Recall the ranges of the integer-valued indices $0<j\leq k<T$ and $0\leq \hat\jmath,\hat k<T$. We denote $\bar{\mc A}_-^k = \bar{\mc A}_-^{k,\d} \, T_\d$, etc.\label{ZTfeyn}}
\end{figure}

\begin{figure}[H]
\centering
\raisebox{-0.5\height}{\includegraphics[scale=0.16]{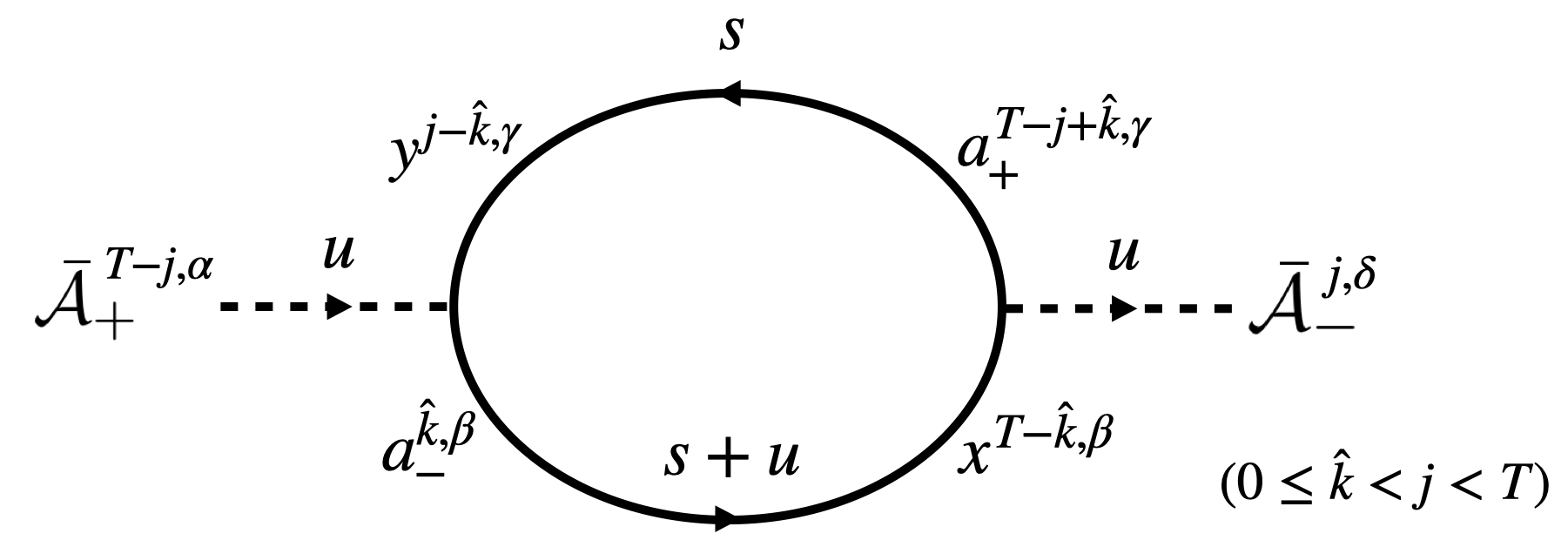}}
\caption[] {\small The only class of divergent diagrams contributing to the 1-loop effective action \rf{univT} for theories with Lax connection of the form \rf{ZTL}.  These diagrams are summed over the indices $j,\hat k$ in the range indicated. \label{ZTdiag}}
\end{figure}

The Feynman rules resulting from \rf{ZTpa} are shown in Figure \ref{ZTfeyn}. 
There is only one type of divergent 1-loop diagram (with a sum over internal fields) shown in Figure \ref{ZTdiag}, leading to
\begin{align}
&2\widehat{S}^{(1)} = \Big(\int \frac{d^2 s}{(2\pi)^2} \frac{1}{s_-(s+u)_+} \Big) \int d^2 \xi \sum_{0\leq {\hat k} < j <T}  X^{\hat k, j}_{+-} \ , \la{expT}\\
&\qquad X^{\hat k, j}_{+-} \equiv \s^{\hat k(\hat k-j)} \,  f^{\g^{{\hat k} -j}} {}_{\a^{-j}\b^{\hat k}} \,  f^{\b^{\hat k}}{}_{\g^{{\hat k} -j} \d^j} \, \bar{\mc A}_+^{\a^{T-j}} \bar{\mc A}_-^{\d^{j}} \ , \la{divT}
\end{align}
where we have dropped finite contributions and  introduced the symbol
\be \la{sdef}
\s = \begin{cases}+1 \  &,\qquad  G\text{ a group} \ , \\
-1 &,\qquad  G\text{ a supergroup} \ ,
\end{cases} 
\ee
to treat the group and supergroup cases at once. We have also introduced the shorthand $\a^j$ for algebra indices only running over the $j$th subspace $\mathfrak{g}_j$. The factor $\s^{\hat k(\hat k-j)}$ in \rf{divT} is just the usual minus sign resulting from fermion loops (since when $\hat k$ and $\hat k-j$ are both odd there is a fermion loop in the supergroup case).

In appendix \ref{Apeq} we show that 
\begin{align}
\sum_{0\leq\hat k<j<T} X^{\hat k, j}_{+-}   =- \frac{2\cG}{T}  \ \Tr[ \bar{\mc A}_+\Big( \sum_{0<j<T}j P_j \Big) \bar{\mc A}_-]  \ .  \la{Xs}
\end{align}
Substituting \rf{Xs} into \rf{divT} and evaluating the divergent part of the loop integral over $s$ as $-\tfrac{1}{2\pi} \log{\frac{\Lambda}{\mu}}$, we obtain the following result
\be\begin{aligned}
&\widehat S^{(1)} = \frac{1}{4\pi} \int d^2 \xi \ \widehat \L^{(1)} \\
&\ddt \L^{(1)} = -\frac{2\cG}{T}  \ \Tr[ \bar{\mc A}_+\Big( \sum_{0<j<T}j P_j \Big) \bar{\mc A}_-] \qquad (t\equiv \log \m) \ .  \la{univT}
\end{aligned}\ee
The result \rf{univT} applies to any theory assuming:

(i) a Lax connection of the form \rf{ZTL},

(ii) the Bianchi Completeness Assumption \ref{BCA}, and

(iii) gauge invariance under $G_0$ with the Lax transforming as \rf{Lg}.

\noindent All of the models considered in this section satisfy these assumptions: for Bianchi Completeness, see appendix \ref{A}; gauge invariance is easy to check in each case.

\subsection{Deformed pure-spinor \texorpdfstring{$\mathbb{Z}_T$}{ZT} cosets} \la{ZTd}
The result \rf{univT} immediately implies  1-loop renormalizability of the    pure-spinor  $\mathbb{Z}_T$ coset \rf{ZT} with the coupling running as
\be
\ddt \h= \frac{4\cG}{T} \ .
\ee

Classically integrable $\eta$- and $\l$-deformations were recently  constructed \cite{Hoare} for  pure-spinor $\mathbb{Z}_T$ cosets,\foot{The existence of the models \rf{ZTe},\rf{ZTl} was previously argued (although the Lagrangians not explicitly constructed) in \cite{Vdi}. The $\mathbb{Z}_4$ cases were constructed in \cite{BR} and \cite{BS} for the $\eta$-model and $\l$-model respectively.}${}^{,}$\foot{As for the $\eta$-models on groups, the linear operator $\mc R$ in \rf{ZTe} is assumed to be an antisymmetric solution of the modified classical Yang-Baxter equation \rf{YBE}. Our conventions for the $\l$-models \rf{ZTl} are related to those of \cite{Hoare} by $g\to g^{-1}$ and reversal of the sign of the WZ term.}
\begin{align}
&\L_\eta = \ -\ha \h \, \Tr[ J_+ \, \PP_-^\eta  \frac{1}{1-\mc \eta {\mc R}_g \PP_-^\eta} J_- ] \ , \qquad J_\pm = g^{-1} \del_\pm g \ , \quad g\in G \ ,\la{ZTe}\\
&\qquad \qquad \PP_\pm^\eta  = \sum_{0<k<T} \frac{1+\tet}{1-\tet}\frac{1-\tet^k}{1+\tet^k}\, P_{\mp k} \ , \qquad \tet = \frac{1-c\eta}{1+c\eta} \ ,  \no \\
\no \\
&\L_\l = k \,  \L_{\rm WZW}(g) + k\, \Tr\big[ g^{-1} \del_+ g A_- - \del_- g g^{-1} A_+ + g^{-1} A_+ g A_- - A_+(P_0 + \PP_-^\l)A_- \big]   \ , \la{ZTl}\\
&\qquad \qquad \PP_\pm^\l= \sum_{0<k<T}\l^k P_{\mp k} \ , \qquad \qquad g\in G \ , \quad A_\pm \in \Lie(G)\ . \no
\end{align}
These models admit Lax connections of the same form \rf{ZTL} as the undeformed $\mathbb{Z}_T$ cosets (and satisfy the other assumptions), with  Lax currents\foot{We may think of the  `physical' Lagrangian for the $\l$-model as that obtained from \rf{ZTl} by integrating out the non-dyanamical vector field $A_\pm$. In \rf{LTl} we denote by $A_\pm(g)$ the vector field's resulting value in terms of the physical field $g$.}
\begin{align}
&\mc A^\eta_\pm = \sum_{0<k<T} \frac{2}{\tet^{k/2}+\tet^{-k/2}} \, P_{\mp k} \, \Big(\frac{1}{1\pm \eta\mc R_g \PP_\pm^\eta}  \, J_\pm\Big)  \ , \la{LTe}\\
&\mc A^\l_\pm = \sum_{0<k<T} \l^{-k/2}\,  P_{\mp  k} \, A_\pm(g)       \  . \la{LTl}
\end{align}
Hence the result \rf{univT} applies and, 
substituting the Lax currents \rf{LTe},\rf{LTl},
one finds the following 1-loop divergences
\begin{align}
&\ddt \widehat\L^{(1)}_\eta =  -\frac{2\cG}{T}\, \Tr\Big[ J_+ \, \frac{1}{1-\eta\PP_-^\eta \mc R_g} \, \Big(  \sum_{0<k<T} \frac{4}{(\tet^{k/2}+\tet^{-k/2})^2} \,k P_{k}\Big)  \,  \frac{1}{1-\eta\PP_-^\eta \mc R_g} \, \,J_- \Big]  \ , \la{Ee}\\
&\ddt \widehat\L^{(1)}_\l = -\frac{2\cG}{T} \, \Tr\Big[ A_+(g) \, \Big(\sum_{0<k<T} \l^{-k}\, k P_k\Big)\,  A_+(g)\Big] \ . \la{El}
\end{align}
A short calculation shows that these counterterms can be absorbed into the models' couplings $(h,\eta)$ and $\l$ respectively (the WZ level $k$ does not run as usual),\foot{The $\eta$- and $\l$-deformed models \rf{ZTe},\rf{ZTl} are related, up to analytic continuation, by Poisson-Lie duality  \cite{Hoare}, with the couplings identified as $k=\h/4\eta$, $\l=(1-c\eta)/(1+c\eta)$. As expected, their 1-loop $\b$-functions \rf{1et} and \rf{1la} match under this identification.}
\begin{align}
&\eta\text{-model}\ : \qquad 
\ddt \begin{pmatrix}
{\rm h}\\ \eta
\end{pmatrix} = \begin{pmatrix}
{\rm h}\\ \eta
\end{pmatrix}
\frac{4\cG}{T}\frac{1}{\rm h} (1-c^2\eta^2) \ , \la{1et}\\
&\l\text{-model}\ : \qquad  \ddt k = 0 \ , \qquad  \ddt \l  = - \frac{2\cG}{T}\frac{\l}{k} \ . \la{1la}
\end{align}

The 1-loop renormalizability of these models is a new result (although it was conjectured in \cite{Hoare}). Even for the undeformed model \rf{ZT}, the RG flow has only previously been studied in  particular cases $T=2$ and $\cG=0$; below we shall compare with the known results in those cases.

These models are thus a new class of examples supporting the link between classical integrability and 1-loop renormalizability. Even for the undeformed cosets, there does not seem to be any manifest global symmetry protecting the particular form of \rf{ZT} other than its integrability.

\subsection{Deformed symmetric spaces (\texorpdfstring{$T=2$}{T=2})}
In the $\mathbb{Z}_2$ case, the pure-spinor coset models \rf{ZT} coincide with the standard $G/H$ symmetric space $\s$-models,
\be
\L = -\ha \h \, \Tr[ J_+ \, P_1 \, J_- ] \ . \la{Z2m}
\ee

Specializing the analysis of section \ref{ZTd} to $T=2$, any theory with Lax connection of the following form (and satisfying the assumptions above)
\be
L_\pm = (P_0 + z^{\pm 1} P_1 ) \, \mc A_\pm \ , \la{Z2L}
\ee
has 1-loop on-shell divergences given by a universal formula,
\be
\ddt \widehat\L^{(1)}  =  -\cG \, \Tr[ \mc A_+ \, P_1 \, \mc A_- ] \la{univ2} \ .
\ee

The $\eta$- and $\l$-deformed $G/H$ models ($T=2$ in \rf{ZTe},\rf{ZTl}) \cite{DMV0,HMS0}
\begin{align}
&\L_\eta = -\ha \h \, \Tr[ J_+ \ P_1 \frac{1}{1-\eta R_g P_1} \  J_- ]  \la{eta2} \\
&\L_\l = k \,  \L_{\rm WZW}(g) +k \, \Tr\big[ g^{-1} \del_+ g A_- - \del_- g g^{-1} A_+  \\
&\hspace{5cm}+ g^{-1} A_+ g A_- - A_+(P_0 + \l^{-1} P_1 )A_- \big]  \ ,\no
\end{align}
are known to be 1-loop renormalizable \cite{SST,Itsios:2014lca,CLa,ah}. Indeed, the formula \rf{univ2} with the Lax currents \rf{LTe},\rf{LTl} (i.e. $T=2$ in  \rf{1et},\rf{1la}) reproduces the correct 1-loop $\b$-functions,
\begin{align}
&\eta\text{-model}\ : \qquad 
\ddt \begin{pmatrix}
{\rm h}\\ \eta
\end{pmatrix} = \begin{pmatrix}
{\rm h}\\ \eta
\end{pmatrix}
\frac{2 \cG}{\rm h} (1-c^2\eta^2) \ , \la{et2}\\
&\l\text{-model}\ : \qquad  \ddt k = 0 \ , \qquad  \ddt \l  = -\cG \frac{\l}{k} \ .\la{la2}
\end{align}

\subsection{Scale invariance with vanishing Killing form (\texorpdfstring{$\cG=0$}{cG=0})}
Now let us fix $G$ to be a supergroup with vanishing   Killing form  ($\cG=0$).
Then the result \rf{univT} implies that any model with Lax connection of the form \rf{ZTL} (and satisfying the other assumptions) is 1-loop scale-invariant,
\be
\ddt \widehat{\L}^{(1)} \Big|_{\cG=0} = 0  \ .  \la{rev}
\ee

In particular, the pure-spinor $\mathbb{Z}_T$ supercosets \rf{ZT} and their $\eta$- and $\l$-deformations \rf{ZTe},\rf{ZTl} are 1-loop scale-invariant when $\cG=0$.
For the undeformed case, this result was already known \cite{KY};
for the $\eta$- and $\l$-deformed models, it is a new result.\foot{This result is of course expected, at least for the $T=4$ case, since the corresponding $\eta$- and $\l$-deformations of the Green-Schwarz type $\mathbb{Z}_4$ supercosets are already known to be 1-loop scale-invariant \cite{eta,Arut,HMS,ah} (see section \ref{out} below).}

One can also consider \textit{multi-parameter} deformations (to appear in \cite{HLS}) for `permutation' $\mathbb{Z}_4$ supercosets of the form $G\times G/G_0^{\rm diag}$. The result \rf{rev} immediately implies that such models are 1-loop scale-invariant, since their Lax connections will take the same form \rf{ZTL}.

\section{General models with simple poles \la{gsim}}
Now let us consider Lax connections with poles  in arbitrary  locations,\foot{One may assume that $\infty$ is not a pole  in \rf{gl} by redefining the spectral parameter. }
\be
L_+ =  \sum_{z_i \in P_+}  \frac{1}{z-z_i} \,  \mc A_+^{(i)} \ , \qquad L_- =  \sum_{w_j \in P_-} \frac{1}{z-w_j} \, \mc A_-^{(j)} 
 \ . \la{gl}
\ee
where $^{(i),(j)}$ should be viewed as flavour indices.\foot{In particular, we emphasize the difference between the flavour indices $^{(i),(j)}$ here and the $\mathbb{Z}_T$ grading indices $^{i,j}$ of section \ref{CD}.}

This is a generalization of the PCM case \rf{fc}, whose poles are $P_\pm = \left\{ \mp 1 \right\}$. In fact \rf{gl} is the most general analytic structure with the following properties:
\begin{enumerate}[(i)]
\item  {\bf There are only simple poles. }

This is just for convenience; it would be easy to relax in principle. 
\item {\bf $L_+$ and $L_-$ do not share any poles ($P_+ \cap P_- = \varnothing$).}

This prevents non-dynamical equations without  derivatives appearing in the zero-curvature equations, excluding cases like the $\ka$-symmetric Green-Schwarz type cosets \cite{GS,Hoare}, which should be treated separately.
\item {\bf There are no constant $\O(z^0)$ terms.}

This assumption is related to the absence of gauge symmetry, since a constant term $L_\pm = \mc A_\pm + \ldots$ would lead to an apparent gauge symmetry under standard gauge transformations of $\mc A$ (cf.\ \rf{Lg}). In principle this can be relaxed but would require a more complicated treatment.

\end{enumerate}
Here we will also assume the Bianchi Completeness Assumption \ref{BCA} in a strong form where the currents $\mc A_\pm^{(i)}$ in \rf{gl} are required to be valued in the full algebra $\Lie(G)$.\foot{For comparison, we only assumed Bianchi Completeness in a weaker form for the $\mathbb{Z}_T$ models in Section \ref{CD}, since we allowed the Lax currents there to be valued in subspaces $\mathfrak{g}_i$ of the algebra (see discussion below \rf{ZT}).}

We will then show that the 1-loop on-shell divergences are given by a universal formula in terms of the Lax currents,
\be
\ddt \widehat \L^{(1)} = 2 \, \cG \sum_{z_i\in P_+, w_j\in P_-} \frac{1}{(z_i-w_j)^2}  \,  \Tr[ \bar{\mc A}_+^{(i)} \bar{\mc A}_-^{(j)}] \ . \la{uniN}
\ee
This formula closely resembles that of the PCM and related models \rf{1p}, dressed with $i,j$ indices (and they match in the case 
$P_\pm = \left\{ \mp 1 \right\}$, 
 $ \mc A_\pm^{(\mp 1)} = \pm \mc A_\pm$).

We shall derive the result \rf{uniN} starting from the general  path integral argument of section \ref{gen}, which applies now with $a_\pm=(a_+^{(i)}, a_-^{(j)})$ denoting the fluctuations of $\mc A_+^{(i)}$, $\mc A_-^{(j)}$.
The zero-curvature equations following from \rf{gl} are
\begin{align}
& \del_+ \mc A_-^{(j)} - \sum_{z_i\in P_+} \frac{1}{z_i-w_j}[\mc A_+^{(i)}\, \mc A_-^{(j)}] = 0 \ , \qquad \del_- \mc A_+^{(i)} - \sum_{w_j\in P_-} \frac{1}{z_i-w_j}[\mc A_+^{(i)}\, \mc A_-^{(j)}] = 0 \ .
\end{align}
Thus choosing a basis $U = (x_j, y_i)$  of scalar fields valued in $\Lie(G)$, the path integral \rf{res} becomes
 \begin{align}
&2 \widehat S^{(1)}(\bar{\mc A}) = - i \log \int \mathcal D x_i \, \mathcal D y_j \, \mathcal D a_+^{(i)} \, \mathcal D a_-^{(j)} \  \la{resN} \\
&\qquad  \qquad\qquad    \exp{ i \int \Tr \Big[  x_j \, \del_+ a_-^{(j)} + y_i \,  \del_- a_+^{(i)}- \tfrac{1}{z_i-w_j} (x_j +y_i) \big( [\bar{\mc A}_+^{(i)},  \, a_-^{(j)} ] +[a_+^{(i)},  \, \bar{\mc A}_-^{(j)} ] \big)\Big] }   \ , \no
\end{align}
where we sum over repeated $i,j$ indices.  
\begin{figure}[H]
\centering
\raisebox{-0.5\height}{\includegraphics[scale=0.12]{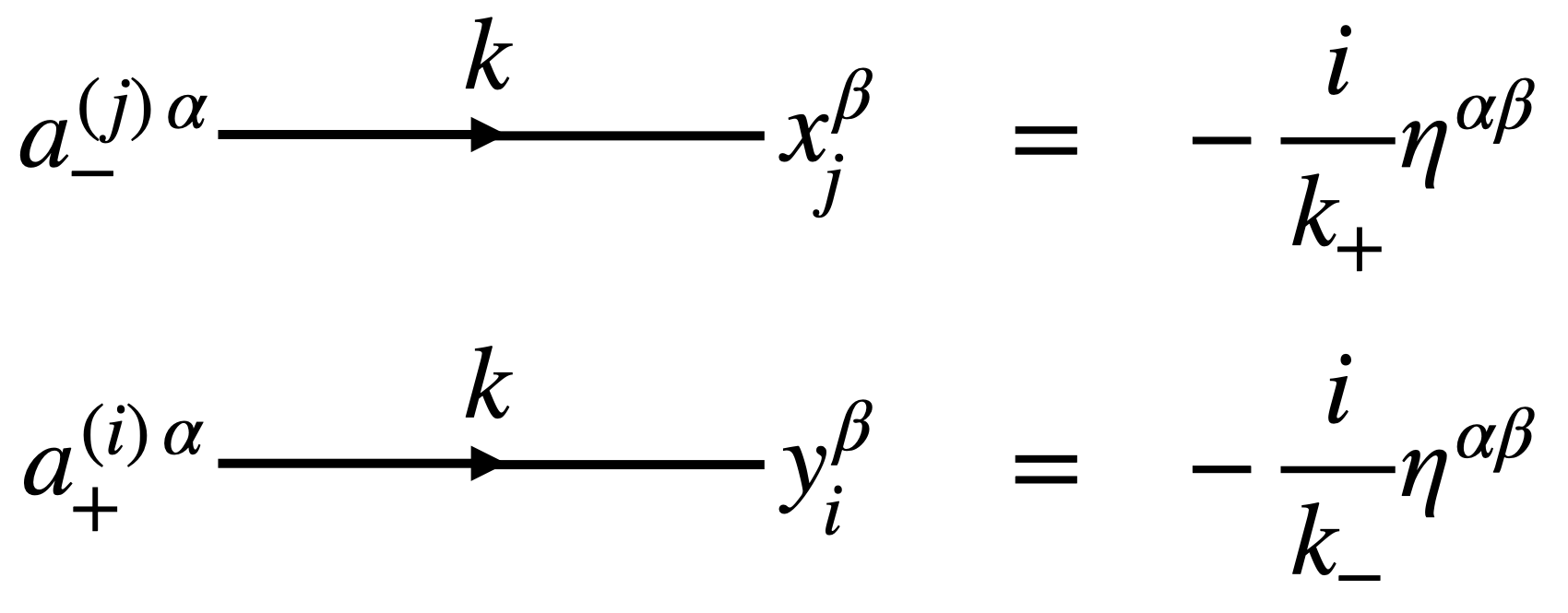}} 
\raisebox{-0.5\height}{\includegraphics[scale=0.12]{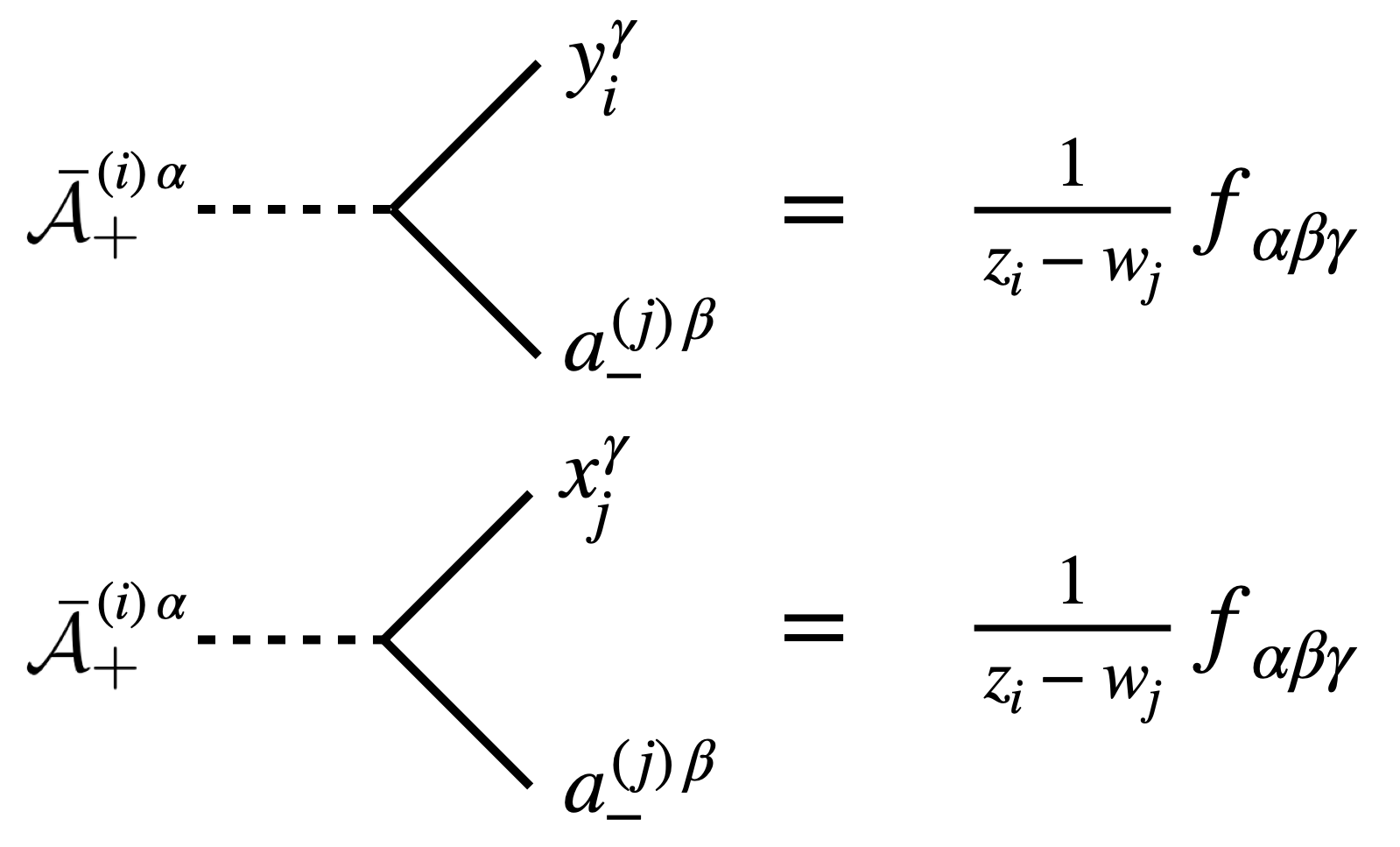}\hspace{1.7cm}\includegraphics[scale=0.12]{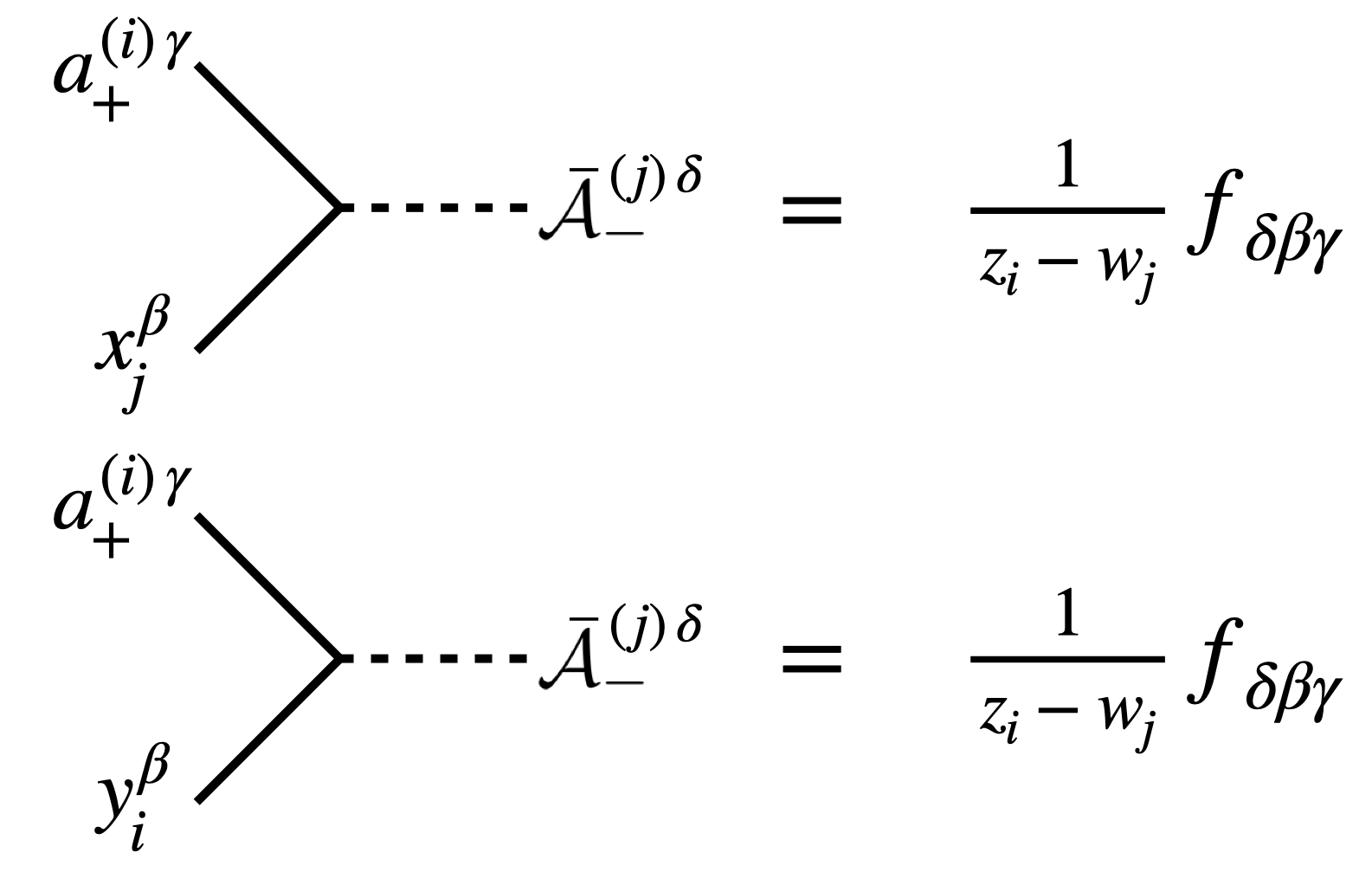}} 
\caption[] {\small The Feynman rules following from the path integral \rf{resN}. The indices $i$ and $j$ respectively run over the sets $P_+$ and $P_-$ in \rf{gl}.
   We denote $\bar{\mc A}_-^{(j)} = \bar{\mc A}_-^{(j)\, \d}\,  T_\d$, etc.
   \label{GNfeyn}}
\end{figure}

\begin{figure}[H]
\centering
\raisebox{-0.5\height}{\includegraphics[scale=0.16]{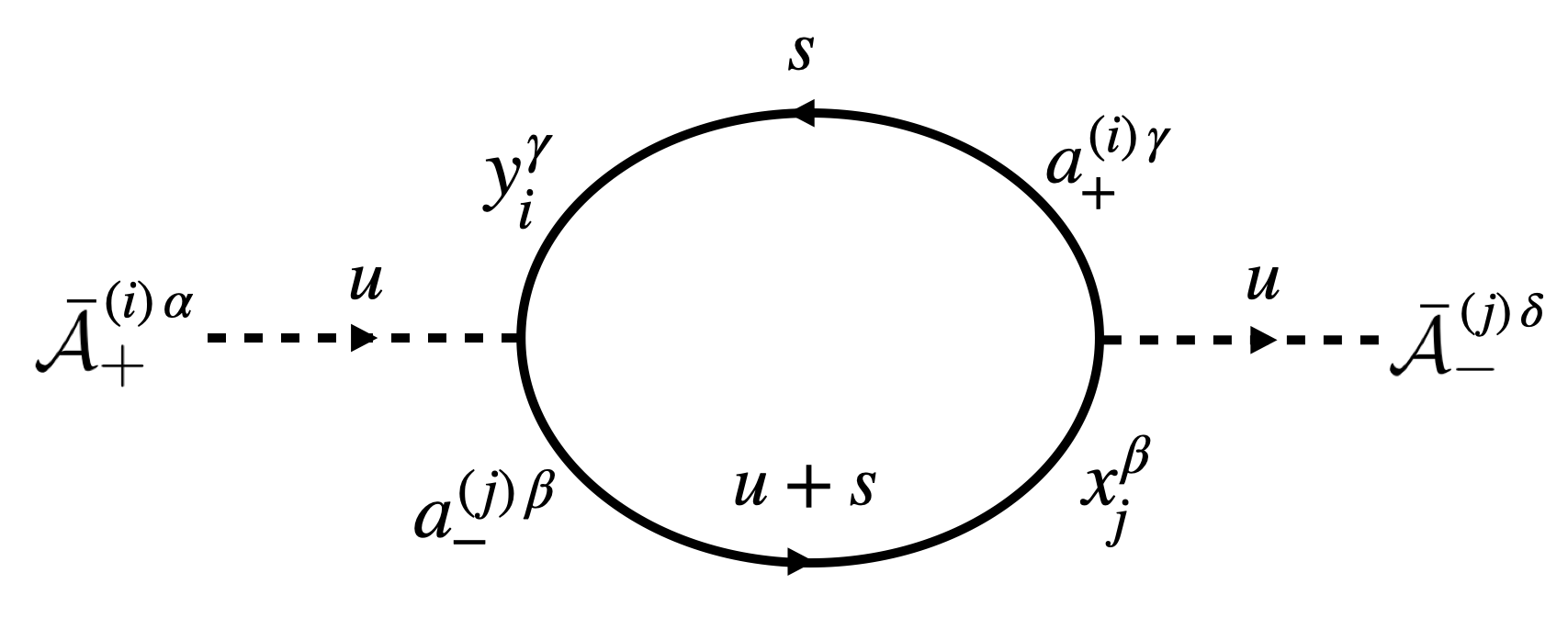}}
\caption[] {\small The only class of divergent diagrams contributing to the 1-loop effective action \rf{divN} for theories with Lax connection of the form \rf{gl}. These diagrams are summed over the indices $i,j$ labelling the external background fields. \label{GNdiag}}
\end{figure}

The  Feynman rules obtained from \rf{resN} are shown in Figure \ref{GNfeyn} and there is only one divergent 1-loop diagram, shown in Figure \ref{GNdiag}, leading to
\be
2 \widehat{S}^{(1)} = \Big(\int \frac{d^2 s}{(2\pi)^2} \frac{1}{s_-(s+u)_+} \Big) \frac{-1}{(z_i-w_j)^2}\, f_{\a\b\g}f_\d{}^{\b\g}\int d^2\xi \, \bar{\mc A}_+^{(i)\, \a}\bar{\mc A}_-^{(j)\, \d}\ . \la{divN}
\ee
The loop integral over $s$ diverges as $-\tfrac{1}{2\pi}\log\tfrac{\Lambda}{\mu}$. Using the identity $f_{\a\b\g}f_\d{}^{\b\g} = -2\cG \eta_{\a\d}$, eq.\ \rf{divN} indeed gives the result \rf{uniN}, where we define $\widehat{S}^{(1)}=\tfrac{1}{4\pi}\int d^2 \xi \, \widehat{\L}^{(1)}$ and $t = \log\mu$.

\subsection{Affine Gaudin models and \texorpdfstring{$G\times G$}{GxG} models\la{GN}}
The 
 classically integrable $\s$-models 
 derived from affine Gaudin models \cite{Vdi,unif} have Lax connections of the form \rf{gl}.\foot{Here we  exclude cyclotomic affine Gaudin models  related to $\mathbb{Z}_T$ gradings \cite{Vdi}  --- cf.\ section \ref{CD}. We also assume the twist-function
 to have only simple zeros (as is conventional \cite{unif}).} 
Thus the result \rf{uniN} applies to all such models satisfying Bianchi Completeness.  It is notable in the affine Gaudin language that the 1-loop divergences  take this simple form, only depending on the poles of the Lax connection (and not the poles of the `twist-function').

The affine Gaudin models' $\b$-functions have only been systematically studied for a certain subclass \cite{DLSS,Falk} with at most double poles in the twist-function.
  It would be good to check our result \rf{uniN} against those works, where the RG flow was interpreted as a flow of the twist-function.
  Prototypical examples of this subclass  are the coupled $G^N$ models \cite{DLMV} and their $\eta$/$\lambda$-deformations \cite{BL},\foot{See also \cite{GSf} and refs.\ therein for particular cases of the $\l$-deformed coupled models that appeared earlier.} which indeed satisfy the Bianchi Completeness Assumption \ref{BCA} due to the sufficient conditions proved in appendix \ref{A}.
 Moreover, our result \rf{uniN} stands to generalize those results,
applying more generally for any pole structure of the twist-function --- as well as any other models not related to affine Gaudin models.

Let us focus on the simplest case\foot{We postpone a full comparison against \cite{DLSS,Falk} as it is somewhat labour-intensive to translate between their formulation as a flow of the twist-function and our own.} of $G\times G$ models without $\eta$/$\lambda$-deformations (here $i,j=1,2$),
\be
\L = -\ha \, \rho_{ij} \, \Tr[ J_+^{(i)} J_-^{(j)} ] + k_i \, \L_{\rm WZ}(g^{(i)}) \ , \qquad J_\pm^{(i)} = (g^{(i)})^{-1} \del_\pm g^{(i)} \ , \quad g^{(1)},g^{(2)}\in G \ . \la{GGmod}
\ee
These theories are classically integrable when the couplings $\rho_{ij}$, $k_i$ solve a certain homogeneous cubic polynomial equation \cite{DLMV}. 
In that case the Lax connection takes the form \rf{gl} with $L_+$ and $L_-$ each having two simple poles at positions depending on the couplings.
Then a short computation shows that the result \rf{uniN} precisely matches the known 1-loop $\b$-functions for the theory \rf{GGmod} as stated in \cite{LT} (and previously computed in \cite{DLSS}).

\section{Discussion \la{Disc}}
We have observed that, since 1-loop divergences can be computed from the classical equations of motion, and 
integrable models have equations of motion of a canonical zero-curvature form, then their 1-loop divergences should take a canonical form. Different classes of theories are distinguished by the different choices of poles for the Lax connection. For each class, we expect the 1-loop $\b$-function to take a universal form in terms of the Lax currents (the residues of the poles). 

This may be seen as a 1-loop quantization scheme that makes integrability manifest. Indeed, we have found that classical integrability places strong constraints on the possible 1-loop divergences, and it would be natural to interpret these as following from Ward identities associated to hidden integrable symmetries.

This approach relies on the fact that on-shell 1-loop divergences are insensitive to the distinction between equations of motion and Bianchi identities --- thus different theories realising the same Lax connection effectively have the same equations of motion. We gave a general path integral argument for this fact and the universality of the 1-loop $\b$-functions in section \ref{gen}. That argument is valid assuming the Bianchi Completeness Assumption \ref{BCA} --- roughly stating that all Bianchi identities needed to relate the Lax currents to the physical fields are included in the zero-curvature equations ---
 which seems to single out $\s$-models.
 It would also be interesting to try to
 incorporate other integrable theories, like sine-Gordon, that are currently excluded by the Bianchi Completeness Assumption.

In practice, our approach renders the computation of 1-loop divergences for integrable $\s$-models much easier. Not only can the divergences of deformed models be obtained from the undeformed ones, but even computing the undeformed models' divergences seems much easier in this formulation. For example, for the $\mathbb{Z}_T$ cosets in section \ref{CD} and the $G^N$ models in section \ref{gsim}, working directly with the Lax currents made the computation much simpler than the alternative of working with the physical fields and directly computing the (generalized) Ricci tensor \rf{beta}.

Let us emphasize that the arguments and conclusions of this paper apply only at 1-loop order. From 2-loops it is unlikely that there could be such a simple universality, since there will be multiple topologies for 2-loop diagrams whose relative contributions will be theory-dependent (see also \cite{HLT,LT} for discussion of higher-loop RG flow in integrable $\s$-models).

\bigskip
In section \ref{PCS} we made extensive checks of our results in the case where $L_+$ and $L_-$ each have one simple pole. Models of this type include the principal chiral model (PCM) with and without WZ term, $\eta$/$\l$-deformations, the NATD of the PCM, and the pseudo-dual of the PCM. In each case we found complete agreement with the known 1-loop $\b$-functions. 

In section \ref{gsim}, we considered the most general classically integrable theory without features associated with gauge symmetries or $\ka$-symmetries (and assuming for convenience only simple poles in the Lax connection). Applying the general path integral argument followed by an explicit 1-loop computation, we found a universal form \rf{uniN} for the 1-loop on-shell divergences.  In section \ref{GN} we considered the consequences of this result for affine Gaudin-derived theories, and checked it expicitly against the  known 1-loop $\b$-functions for integrable $G\times G$ models.

\bigskip
For models with Lax connection \rf{ZTL} of the same form as pure-spinor $\mathbb{Z}_T$ cosets, we argued that the path integral argument still applies despite two complications. First, the Lax currents are constrained to subspaces of the full algebra: this can always be accommodated in the argument. Second, the zero-curvature equations have an apparent gauge invariance: this can be accommodated assuming the theory has a genuine gauge symmetry under which the Lax transforms as a gauge field.

By an explicit computation, we derived a universal formula  \rf{univT} for the 1-loop on-shell divergences. This led to a new result for pure-spinor $\mathbb{Z}_T$ cosets and their $\eta$/$\l$-deformations: these models are 1-loop renormalizable, with the $\b$-functions \rf{1et},\rf{1la}.
For supergroups with vanishing Killing form ($\cG=0$), it follows that all models with this type of Lax connection are scale-invariant, including  $\eta$/$\lambda$-deformations of pure-spinor supercosets and
multi-parameter deformations to appear in \cite{HLS}.

Establishing these models'
 scale invariance 
 is a first indication that they may define consistent string theories.\foot{In the pure-spinor case, one further needs conformal invariance and BRST invariance when the theory is coupled to ghosts \cite{PSC}.}
 Remarkably, this follows just from their classical integrable structure.

\section{Outlook \la{out}}
The 1-loop scale invariance of the deformed pure-spinor $\mathbb{Z}_T$ supercoset models with $\cG=0$ is natural, since the corresponding deformed $\mathbb{Z}_4$ supercosets of Green-Schwarz type \cite{eta,HMS} are already known to be scale-invariant \cite{eta,Arut,HMS,ah}. In fact, those Green-Schwarz models are actually 2d Weyl-invariant when equipped with certain dilatons (assuming a unimodularity condition on the R-matrix in the $\eta$-deformed case, and vanishing total central charge) \cite{BW,unimod}. In combination with the correct $\ka$-invariance, this implies that the target spaces are supergravity solutions. This is a remarkable and surprising fact: these deformations were originally written down with the purpose of preserving classical integrability, and miraculously, in some cases, they are exactly marginal.

The present work suggests a possible explanation: 
perhaps the 1-loop Weyl anomaly --- like the scale anomaly --- takes a universal form 
for the undeformed and deformed models,
 due to their Lax connections having the same analytic structure.
Thus Weyl invariance of the undeformed model may, under some conditions, imply Weyl invariance of the deformed one. In this way one may propose to find new supergravity solutions using 2d integrability, by finding new realizations of a Lax connection with a given analytic structure.

For this idea to make sense,\foot{In principle the Green-Schwarz models also have the added complication of fermionic $\ka$-symmetry. However, this should not pose a problem as it may be fixed in a `universal' way, similar to the handling of the gauge symmetry in eq.\ \rf{Lg}.} one needs to demonstrate a `universal' way of computing the Weyl anomaly, which generally differs from the scale anomaly $\ddt {\widehat \L}^{(1)}$ by a  diffeomorphism-type term $L_V(G+B)_{mn} \del_+ x^m \del_- x^n$.
For 1-loop Weyl invariance, 
the scale anomaly must vanish and the  vector $V$ must take the dilaton form $V_m = \del_m \phi$ (which can be cancelled by a local counterterm). 
We expect that there would generically be some obstruction to the universality of the Weyl anomaly, which would introduce non-universal vectors $V$ of  non-dilaton form and break Weyl invariance. Demanding the absence of such terms may lead to the unimodularity condition in the $\eta$-deformed case. One may then hope to use this approach to explain the origin of the dilatons \cite{BW} for these backgrounds from first principles.

\bigskip

An initial motivation for this study was to prove the 1-loop RG stablity of the space of classically integrable $\s$-models. For a large classes of theories, we have carried out a first step:  computing the divergences in a universal way in terms of the classical integrable data. The next step is to prove that these divergences can be absorbed into the Lax connection\foot{In many cases a divergence is also absorbed into an overall `radius' coupling that rescales the action.} --- i.e. that the 1-loop effective equations of motion (from varying the effective action) are of Lax form. Indeed, this is what we see in examples: in all the models mentioned in this paper, the 1-loop divergences can be absorbed into the classical coupling constants in the Lax connection. In order to prove this statement, it would probably be necessary to use the assumption that the  Lax connection actually corresponds to a Lagrangian theory, which places some constraints on the possible Lax currents. It would be interesting to address this problem in the future.

Another approach would be to \textit{assume} that classically integrable models are 1-loop renormalizable and use the 1-loop divergences computed in this paper to try to bootstrap the Lagrangians of classically integrable theories.

\bigskip

Many  integrable  2d field theories can be obtained, at the classical level, from 4d Chern-Simons theory localized to 2d defects \cite{CWY}. In order to understand whether this relation still holds at the quantum level, it would be desirable to correctly compute some 2d quantum observables, e.g.\ 1-loop $\b$-functions, directly from the 4d theory.

We remark that our approach is reminiscent of the 4d Chern-Simons theory, since the fundamental object of the path integral \rf{res} is the Lax connection and its fluctuations. Similarly, in 4d Chern-Simons, the 2d Lax connection is essentially the fundamental object (it is obtained from the  4d gauge field by fixing a particular gauge and solving some equations of motion\foot{Moreover, in our approach, solving some equations of motion before quantizing should be acceptable, since they are interchangable with Bianchi identities at the level of 1-loop divergences.}). Thus one may hope to prove a 4d-2d equivalence of 1-loop divergences using a similar approach.

\bigskip
\section*{Acknowledgments}
I am grateful to B.~Hoare,  S.~Lacroix and A.~A.~Tseytlin for comments on this manuscript and to C.~Bachas,  M.~F.~Paulos, D.~Thompson, and B.~Vicedo for stimulating discussions. This work was supported by  the Institut Philippe Meyer at the \'{E}cole Normale Sup\'{e}rieure in Paris.

\bigskip

\appendix
\addtocontents{toc}{\protect\setcounter{tocdepth}{1}}

\section{On the Bianchi Completeness Assumption} \label{A}
\def\theequation{A.\arabic{equation}}
\setcounter{equation}{0}
The general argument of section \ref{gen}  relies on the assumption \ref{BCA} of `Bianchi completeness'. The assumption is that, to leading order in the background field expansion \rf{ax}, a subset of the zero-curvature equations (which we  call the `Bianchi identities') may be solved to express the Lax currents in terms of the physical fields. In this appendix we  derive sufficient conditions for Bianchi completeness, and show that all  the examples in  this article are Bianchi complete.

\bigskip
For 
almost all
examples considered in this article, the Bianchi identities take the form $F_\pm(g^{-1} \del g) \equiv 0$, where the Lax currents are related by $\mc A_\pm (g) = O_\pm(g) \, g^{-1} \del_\pm g$, the physical field $g$ belongs to some (super)group $\hat G$ and the operators $O_\pm(g):\Lie(\hat G)\to\Lie(\hat G)$ are invertible for all $g$ (away from singular points in the target space geometry).\foot{For the $\mathbb{Z}_T$ coset models of section \ref{CD}, $\mc{A}_\pm$ here is the sum of the Lax currents $P_{\mp i} \mc A_\pm$ in \rf{ZTL}; for the general simple-pole models of section \ref{gsim}, $\mc A_\pm$ is a direct sum of the Lax currents $\mc A_+^{(i)}, \mc A_-^{(j)}$. This does not affect the following analysis. Note that in general the field space $\hat G$ need not be the same as the group $G$ (where the Lax is valued in $\Lie(G)$).} We shall show that this is a sufficient condition for Bianchi completeness.

By definition, the physical form of the Lax current, $\mc A_\pm = O_\pm(g) \, g^{-1} \del_\pm g$, 
solves the Bianchi identity,
\be
0 \equiv F(O_\pm(g)^{-1} \mc A_\pm) = O_-(g)^{-1} \del_+\mc A_- - O_+(g)^{-1} \del_-\mc A_+ +\O(\del g  \, \mc A) + \O(\mc A^2)  \ . \la{Ba}
\ee
Thus expanding $g=\bar g(1+\phi)$ around a background,  the corresponding fluctuation of $\mc A_\pm(g)$,
\be
a_\pm(\phi) = \big[O_\pm(\bar g) \del_\pm + N_\pm (\bar g)\big] \phi \la{lph} \ ,
\ee
is a solution of the Bianchi identity \rf{Ba} in its linearized form,
\be
\big[ O_-(\bar g)^{-1} \del_++ M_+(\bar g) \big] a_-  - \big[ O_+(\bar g)^{-1} \del_-  + M_-(\bar g) \big] a_+  \equiv 0\ . \la{lB}
\ee
Here $M_\pm(\bar g),N_\pm(\bar g)$ denote some linear operators $\Lie(\hat G) \to \Lie(\hat G)$ whose exact form is not important. We have used the fact the background field is on-shell to replace $g\to\bar g$ in \rf{lB} (see comment below eq.\ \rf{new}).
 
For Bianchi completeness, we must prove the converse: that \rf{lph} is the most general solution of \rf{lB}. Consider a general solution $a_\pm$ of \rf{lB}. One may always write $a_+ = \big[O_+ \del_+ + N_+\big] \phi_0 $ where\foot{In \rf{p0} and below, we denote $\int d\xi^+ =\int_{-\infty}^{\xi^+} dx^+$, etc. Here we use $-\infty$ as a basepoint, but in fact it could be any fixed value (e.g.\ for a theory on a cylinder rather than a plane).}
\be
\phi_0 = \big( \mc P \exp \int d\xi^+\, O_+^{-1}N_+ \big)^{-1} \, \int d\xi^+ \,  \big( \mc P \exp \int d\xi^+\, O_+^{-1}N_+ \big)  \  O_+^{-1} a_+ \ , \la{p0}
\ee
where, for convenience, we are suppressing the dependence on $\bar g$. Then subtracting $a_\pm(\phi_0)$ from $a_\pm$, we obtain a solution $\a_\pm$ to \rf{lB} with $\a_+=0$,
\be
\big[ O_-^{-1} \del_+ + M_+ \big] \a_- = 0 \ , \qquad \a_+=0  \qquad \qquad (\a_\pm = a_\pm - a_\pm(\phi_0) ) \ . \la{ale}
\ee
The most general solution to \rf{ale} is
\be
\a_- = \big(\mc P \exp \int d\xi^+\, O_- M_+ \big)^{-1}\b_-(\xi^-)\ , \qquad \a_+ = 0 \ . \la{ceq}
\ee
i.e.\ they are classified by their $\xi^+ \to -\infty$ behaviour,
\be
\a_- \sim \b_-(\xi^-) \ , \la{beh1}
\ee
where here $\b_-(\xi^-) \in \Lie(G)$ is an arbitrary function of $\xi^-$.

One class of solutions of \rf{ale} are of the `physical' form \rf{lph},
\begin{align}
\a_\pm=a_\pm(\phi_1) \ , \qquad &a_+(\phi_1)=\big[O_+ \del_+ + N_+\big] \phi_1 = 0 \ , \la{alps}\\
&\phi_1=\big(\mc P \exp \int d\xi^+\, O_+^{-1}N_+\big)^{-1} Z(\xi^-) \ , \la{p1}
\end{align}
where $Z(\xi^-)$ is an arbitrary function. Substituting \rf{p1} into \rf{alps}, these physical solutions indeed take the form
\be
\a_+ = 0 \ ,\qquad \a_- = \big[O_-\del_- + N_-\big] (\mc P \exp \int d\xi^+\, O_+^{-1}N_+ \big)^{-1} Z(\xi^-) \ . \la{pso}
\ee
In particular, these must be a subset of the general solutions \rf{ceq}. In the limit $\xi^+ \to -\infty$, the physical solutions \rf{pso} behave as
\be
\a_- \sim \big[O_-\del_- + N_- \big]\Big|_{\xi^+=-\infty} Z(\xi^-) \ . \la{beh2}
\ee
Thus, in fact, these must be \textit{all} of the solutions \rf{ceq} since, equating \rf{beh1} and \rf{beh2}, we can always solve for 
\be 
Z(\xi^-) = \big( \mc P \exp \int d\xi^-\, O_-^{-1} N_-\big)^{-1} \, \int d\xi^- \, \big( \mc P \exp \int d\xi^- \, O_-^{-1}N_-\big) \ O_-^{-1}\b_-(\xi^-)\Big|_{\xi^+=-\infty}  \ .
\ee
Hence the most general solution to \rf{ale} is $\a_\pm = a_\pm(\phi_1)$. Then the most general solution to \rf{lB} is of the required form $a_\pm = a_\pm(\phi_0+\phi_1)$ (with $\phi_0,\phi_1$  given by \rf{p0},\rf{p1}).

\bigskip
Another sufficient condition is obtained from the one above by zooming in on the identity element of the group $\hat G$ to obtain an algebra as the field space: the Bianchi identities take the form $\del_+ (\del_- v ) \equiv \del_- (\del_+ v)$ where $v \in \Lie(\hat G)$ is the physical field and the Lax currents are related by $\mc A_\pm = O_\pm(\bar v) \, \del_\pm v$ for invertible operators $O_\pm$ depending on $\bar v$.
Equations \rf{lph} and \rf{lB} are unchanged in that limit (with $\bar g \to \bar v$) and the argument above goes through.
 The only remaining cases in this article, the NATD \rf{NATD} and the pseudo-dual \rf{pseudo} of the PCM, are of this type.

\section{Conventions and identities for (super)algebras} \label{Apeq}
\def\theequation{B.\arabic{equation}}
\setcounter{equation}{0}
\subsection{Conventions}
In this paper all Lie (super)groups are assumed to be semisimple. 

\medskip \noindent
We label the generators $T_\a$ by indices $\a,\b,\g$ and define the structure constants by \mbox{$[T_\a,T_\b\}=i f^\a{}_{\b\g}$}.\sloppy 

\medskip \noindent
We decompose algebra elements with the notation $X = X^\a T_\a$. 

\medskip \noindent
Indices are lowered and raised  by  $\eta_{\a\b} = \Tr[ T_\a T_\b]$ and its inverse acting on the left, e.g. \mbox{$f_{\a\b\g} = \eta_{\a\d}  f^\d{}_{\b\g} $.}\sloppy   

\medskip \noindent
For supergroups we define $|\b|$ to be an even/odd integer if the generator $T_\b$ is bosonic/fermionic respectively.

\medskip \noindent
The dual Coxeter number of $G$ is denoted $\cG$.

\medskip \noindent
The structure constants satisfy $\s^{|\b|} f^\b{}_{\a\g} f^\g{}_{\d\b} = 2\cG\eta_{\a\d}$, where the symbol $\s$ defined in \rf{sdef} is $+1$ or $-1$ for  groups or supergroups respectively.

\subsection{\texorpdfstring{$\mathbb{Z}_T$}{ZT} grading conventions}
For $\mathbb{Z}_T$-graded (super)abgebras, we denote as $P_j$ the projectors onto the subspaces $\mathfrak{g}_j$ with grading $j=0,1,\ldots,T-1$. 

\medskip \noindent
The grading means that $[\mathfrak{g}_i,\mathfrak{g}_j]\subset \mathfrak{g}_{i+j}$.

\medskip \noindent
We will sometimes denote projections as $\mc A^j \equiv P_j \mc A$ and write $\a^j$ for  algebra indices only running over the generators in $\mathfrak{g}_j$.

\medskip \noindent
For superalgebras we assume the even/odd graded generators to be bosonic/fermionic respectively.

\subsection{Derivation of eq.\ \rf{Xs}}

In section \ref{CD} we evaluated the 1-loop diagram in Figure \ref{ZTdiag}, obtaining the expression \rf{expT} depending on the following quantity (in the context of $\mathbb{Z}_T$-graded (super)algebras)
\be
X^{\hat k, j}_{+-} \equiv \s^{\hat k(\hat k-j)} \,  f^{\g^{{\hat k} -j}} {}_{\a^{-j}\b^{\hat k}} \,  f^{\b^{\hat k}}{}_{\g^{{\hat k} -j} \d^j} \, \bar{\mc A}_+^{\a^{T-j}} \bar{\mc A}_-^{\d^{j}} \ .
\ee
The purpose of this appendix is simplify the sum $\sum_{0\leq\hat k<j<T} X^{\hat k, j}_{+-}$ to the form \rf{Xs}. Here we use hats to distinguish between the ranges $0\leq \hat \jmath, \hat k<T$ and $0< j,k <T$.

Using the graded antisymmetry $f^\a{}_{\b^j\g^k} = \s^{jk} f^\a{}_{\g^k\b^j}$, we rearrange
\begin{align}
X^{\hat k, j}_{+-}  &= \s^{\hat k(\hat k-j)} \s^{j(\hat k-j)} f^{\g^{{\hat k} -j}} {}_{\a^{-j}\b^{\hat k}} \,  f^{\b^{\hat k}}{}_{\d^j \g^{{\hat k} -j} }  \, \bar{\mc A}_+^{\a^{T-j}} \bar{\mc A}_-^{\d^{j}}   \no \\
&= -\s^{\hat k-j}  f^{\g^{{\hat k} -j}} {}_{\a^{-j}\b^{\hat k}} \,  f^{\b^{\hat k}}{}_{\d^j \g^{{\hat k} -j} } \, \bar{\mc A}_+^{\a^{T-j}} \bar{\mc A}_-^{\d^{j}} \no \\
&= -\Tr[ \ad_{\mc A_+^{T-j}} \,  P_{\hat k} \, \ad_{\bar{\mc A}_-^j}] \ , \la{trl}
\end{align}
where $\Tr$ in \rf{trl} is the appropriate (super)trace in the adjoint representation, \hbox{$\Tr[X] = \sum_{\hat \jmath}  \s^{\hat \jmath}\, \tr[  P_{\hat \jmath} X]$}. \sloppy

 Following \cite{KY}, assuming the zero-graded subalgebra $\mathfrak{g}_0$ to be semisimple, then the (graded) Jacobi identity implies
\hbox{$\Tr[ \ad_{\bar{\mc A}_+^{T-j}} \,  P_{\hat k} \, \ad_{\bar{\mc A}_-^j}] =\Tr[ \ad_{\bar{\mc A}_+^{T-j}} \,  P_{\hat k+j} \, \ad_{\bar{\mc A}_-^j}]$.} \sloppy 
Applying this identity $n$ times for each $n=0,\ldots T-1$, we obtain
\be
\sum_{0\leq\hat k<j<T} X^{\hat k, j}_{+-}  = - \frac{1}{T}  \sum_{0<j<T}\ \sum_{0\leq n <T} \ \sum_{nj \leq \hat k < (n+1)j} \Tr[ \ad_{\bar{\mc A}_+^{T-j}} \,  P_{\hat k} \, \ad_{\bar{\mc A}_-^j}] 
\ee
In this sum, each $0\leq \hat k<T$ is hit $j$ times, so re-organising and applying the identity $\Tr[ \ad_X \ad_Y] = 2 \cG \Tr[ XY]$ gives
\begin{align}
\sum_{0\leq\hat k<j<T} X^{\hat k, j}_{+-}  =  -  \sum_{0<j<T}\frac{j}{T}  \ \Tr[ \ad_{\bar{\mc A}_+^{T-j}}  \ad_{\bar{\mc A}_-^j}] =- \frac{2\cG}{T}  \ \Tr[ \bar{\mc A}_+\Big( \sum_{0<j<T}j P_j \Big) \bar{\mc A}_-]  \ ,  \la{XsA}
\end{align}
matching \rf{Xs} as required.

\section{\texorpdfstring{`On-shell'}{'On-shell'} 1-loop duality as a generalization of T-duality \la{T}}
\def\theequation{C.\arabic{equation}}
\setcounter{equation}{0}
In this section, we consider abelian T-duality and show that the same logic (1-loop equivalence between equations of motion and Bianchi identities) leads to the usual statement that T-duality commutes with the 1-loop RG flow.

Consider a $\s$-model on a target space with $U(1)$ isometry,\foot{For simplicity we assume the absence of mixed $\del x \del y$ terms and B-field couplings in \rf{Td}.}
\be
- \L = a(y) (\del x)^2 + g_{mn}(y) \del y^m \del y^n \ . \la{Td}
\ee
The equations of motion are
\be
\del (a(y)  \del x) = 0 \ , \qquad \del_p a(y) (\del x)^2 +  \del_p g_{mn}(y) \del y^m \del y^n  - 2\del(g_{pm}(y) \del y^m) = 0 \ . \la{EOM1}
\ee
These equations can be formulated with a Bianchi identity imposing $A\equiv\del x$ as follows
\be
\del (a(y)  A) = 0 \ , \qquad \del_p a(y) A^2 +  \del_p g_{mn}(y) \del y^m \del y^n  - 2\del(g_{pm}(y) \del y^m) = 0 \ , 
\qquad \e^{ab} \del_a A_b \equiv 0 \ . \la{Teqs}
\ee

The T-dual with respect to $x$ is related to \rf{Td} by  $a(y) \to  a(y)^{-1}$,
\be
- \widetilde \L = \frac{1}{a(y)} (\del \tilde x)^2 + g_{mn}(y) \del y^m \del y^n \ . \la{Td2}
\ee
The equations of motion of the dual theory \rf{Td2} may similarly be fomulated with a Bianchi identity as (putting $a(y)\to a(y)^{-1}$ and $A \to \tilde A$ in \rf{EOM1})
\be \begin{aligned}
\del (a(y)^{-1}  \tilde A) = 0 \ , \qquad  - a(y)^{-2} \del_p a(y) \tilde{A}^2 +  \del_p g_{mn}(y) \del y^m \del y^n  - 2\del(g_{pm}(y) \del y^p) = 0\ , \hspace{0.8cm}&\la{EOM2} \\
\e^{ab} \del_a \tilde A_b \equiv 0& \ .
\end{aligned} \ee
The dual set of equations \rf{EOM2} are the same as the original ones \rf{EOM1} up to (i) swapping the first and last equations and (ii) the redefinition $ A_a = a(y)^{-1} \e_{ab} \tilde A^b$.
According to the argument in section \ref{gen}, swapping the first and last equations (an equation of motion and a Bianchi identity) does not affect the 1-loop on-shell divergences.\foot{Although the argument of section \ref{gen} was formulated for integrable equations originating from zero-curvature equations, we also expect it to apply more generally for non-integrable theories. In the present case, it is easy to see that the argument goes through (with $y^i$ as spectator fields) because the `physical fields' $x,\tilde x$ do not appear explicitly in the equations \rf{EOM1},\rf{EOM2}.}

Since $A_a = \del_a x$,  $\tilde A_a = \del_a \tilde x$ on-shell, then the redefinition $ A_a = a(y)^{-1} \e_{ab} \tilde A^b$ leads to the following relation between their 1-loop on-shell divergences,\foot{Here we omit the hat on the 1-loop effective Lagrangians $\widehat \L^{(1)}$ to simplify notation.} 
\be
 \ddt \widetilde{\L}^{(1)}= \ddt \L^{(1)} \Big|_{ \del_a x = a(y)^{-1} \e_{ab}\del^b \tilde x} \la{rT}
\ee

Let us show that the relation \rf{rT} is the standard one.  One can understand the RG flow of the model \rf{Td} as a flow of the functions $a(y),g_{mn}(y)$ since these are the only freedoms allowed by the global symmetries (neglecting possible diffeomorphism terms, since we consider only on-shell divergences),
\be
- \ddt \L^{(1)} = \frac{d a(y)}{dt} (\del x)^2 + \frac{dg_{mn}(y)}{dt} \del y^m \del y^n \la{1lt} \ .
\ee
The usual path integral relation between the T-dual models \rf{Td},\rf{Td2} implies that the duality commutes with the 1-loop RG flow. This means the dual on-shell effective action is obtained by putting $a(y)\to a(y)^{-1}$ in \rf{1lt},
\be
- \ddt \widetilde{\L}^{(1)} = -a(y)^{-2}  \frac{d a(y)}{dt} (\del \tilde x)^2 + \frac{dg_{mn}(y)}{dt} \del y^m \del y^n \ . \la{2lt} 
\ee
Indeed \rf{1lt},\rf{2lt} are related precisely according to \rf{rT}.

It is natural to understand the relation between T-duality and the RG flow in this way, since the `interpolating model' for the T-dual pair \rf{Td},\rf{Td2},
\be
- \L = a(y) A^2 + g_{mn}(y) \del y^m \del y^n + \tilde x \, \e^{ab} \del_a A_b \ , 
\ee
may be seen as recasting the Bianchi identity $\e^{ab} \del_a A_b \equiv 0$ of \rf{Teqs} as an equation of motion.

\bigskip



\begin{thebibliography}{30}
\bibitem{maillet}
J.~M.~Maillet,
``Kac-Moody Algebra and Extended Yang-Baxter Relations in the O($N$) Nonlinear $\sigma$ Model,''
\doilink{Phys. Lett. B \textbf{162}, 137 (1985)}{doi:10.1016/0370-2693(85)91075-5};
``New Integrable Canonical Structures in Two-dimensional Models,''
\doilink{Nucl. Phys. B \textbf{269}, 54 (1986)}{doi:10.1016/0550-3213(86)90365-2}.

S.~Lacroix, M.~Magro and B.~Vicedo,
``Local charges in involution and hierarchies in integrable sigma-models,''
\doilink{JHEP \textbf{09}, 117 (2017)}{doi:10.1007/JHEP09(2017)117}
\arxivlink{1703.01951}.


\bibitem{friedan} 
G.~Ecker and J.~Honerkamp,
``Application of invariant renormalization to the nonlinear chiral invariant pion lagrangian in the one-loop approximation,''
\doilink{Nucl.\ Phys.\ B {\bf 35}, 481 (1971)}{10.1016/0550-3213(71)90468-8};
J.~Honerkamp,
``Chiral multiloops,''
\doilink{Nucl.\ Phys.\ B {\bf 36}, 130 (1972)}{10.1016/0550-3213(72)90299-4}.

D.~H.~Friedan,
``Nonlinear Models in Two + Epsilon Dimensions,''
\doilink{Annals Phys.\ {\bf 163}, 318 (1985)}{doi:10.1016/0003-4916(85)90384-7};
``Nonlinear Models in Two + Epsilon Dimensions,''
\doilink{Phys.\ Rev.\ Lett.\ {\bf 45}, 1057 (1980)}{10.1103/PhysRevLett.45.1057}.

T.~L.~Curtright and C.~K.~Zachos,
``Geometry, Topology and Supersymmetry in Nonlinear Models,''
\doilink{Phys. Rev. Lett. \textbf{53} (1984), 1799}{
doi:10.1103/PhysRevLett.53.1799}.

%
\bibitem{intRG} 
V.~A.~Fateev, E.~Onofri and A.~B.~Zamolodchikov,
``Integrable deformations of the $O(3)$ sigma model. The sausage model,''
\doilink{Nucl.\ Phys.\ B {\bf 406}, 521 (1993)}{10.1016/0550-3213(93)90001-6}.

V.~A.~Fateev,
``Classical and Quantum Integrable Sigma Models. Ricci Flow, ``Nice Duality'' and Perturbed Rational Conformal Field Theories,''
\doilink{J. Exp. Theor. Phys. \textbf{129}, no.4, 566-590 (2019)}{10.1134/S1063776119100042}
\arxivlink{1902.02811}.

S.~L.~Lukyanov,
``The integrable harmonic map problem versus Ricci flow,''
\doilink{Nucl.\ Phys.\ B {\bf 865}, 308 (2012)}{10.1016/j.nuclphysb.2012.08.002}
\arxivlink{1205.3201}.
%


\bibitem{Squellari}
R.~Squellari,
``Yang-Baxter $\sigma$ model: Quantum aspects,''
\doilink{Nucl. Phys. B \textbf{881} (2014), 502-513}{
doi:10.1016/j.nuclphysb.2014.02.009}
\arxivlink{1401.3197}.

\bibitem{Itsios:2014lca}
G.~Itsios, K.~Sfetsos and K.~Siampos,
``The all-loop non-Abelian Thirring model and its RG flow,''
\doilink{Phys. Lett. B \textbf{733} (2014), 265-269}{
doi:10.1016/j.physletb.2014.04.061}
\arxivlink{1404.3748}.


\bibitem{DLSS}
F.~Delduc, S.~Lacroix, K.~Sfetsos and K.~Siampos,
``RG flows of integrable $\sigma$-models and the twist function,''
\doilink{JHEP \textbf{02} (2021), 065}{
doi:10.1007/JHEP02(2021)065}
\arxivlink{2010.07879}.

\bibitem{LT}
N.~Levine and A.~A.~Tseytlin,
``Integrability vs. RG flow in $G \times G$ and $G \times G /H$ sigma models,''
\doilink{JHEP \textbf{05} (2021), 076}{
doi:10.1007/JHEP05(2021)076}
\arxivlink{2103.10513}.

\bibitem{ah}
C.~Appadu and T.~J.~Hollowood,
``Beta function of $k$ deformed $AdS_5 \times S^5$ string theory,''
\doilink{JHEP {\bf 1511}, 095 (2015)}{10.1007/JHEP11(2015)095},
\arxivlink{1507.05420}.

\bibitem{DLMV}
F.~Delduc, S.~Lacroix, M.~Magro and B.~Vicedo,
``Integrable Coupled $\sigma$ Models,''
\doilink{Phys. Rev. Lett. \textbf{122}, no.4, 041601 (2019)}{doi:10.1103/PhysRevLett.122.041601}
\arxivlink{1811.12316};
``Assembling integrable $\sigma$-models as affine Gaudin models,''
\doilink{JHEP \textbf{06}, 017 (2019)}{doi:10.1007/JHEP06(2019)017}
\arxivlink{1903.00368}.



\bibitem{Berkovits}
N.~Berkovits, M.~Bershadsky, T.~Hauer, S.~Zhukov and B.~Zwiebach,
``Superstring theory on AdS(2) x S**2 as a coset supermanifold,''
\doilink{Nucl. Phys. B \textbf{567} (2000), 61-86}{
doi:10.1016/S0550-3213(99)00683-5}
\arxivlink{hep-th/9907200}.

N.~Berkovits,
``Super Poincare covariant quantization of the superstring,''
\doilink{JHEP \textbf{04} (2000), 018}{
doi:10.1088/1126-6708/2000/04/018}
\arxivlink{hep-th/0001035}.

B.~C.~Vallilo,
``Flat currents in the classical AdS(5) x S**5 pure spinor superstring,''
\doilink{JHEP \textbf{03} (2004), 037}{
doi:10.1088/1126-6708/2004/03/037}
\arxivlink{hep-th/0307018}.


\bibitem{Young}
C.~A.~S.~Young,
``Non-local charges, Z(m) gradings and coset space actions,''
\doilink{Phys. Lett. B \textbf{632} (2006), 559-565}{
doi:10.1016/j.physletb.2005.10.090}
\arxivlink{hep-th/0503008}.

\bibitem{Hoare}
B.~Hoare,
``Integrable deformations of sigma models,''
\doilink{J. Phys. A \textbf{55} (2022) no.9, 093001}{
doi:10.1088/1751-8121/ac4a1e}
\arxivlink{2109.14284}.

\bibitem{HL}
B.~Hoare and S.~Lacroix,
``Yang\textendash{}Baxter deformations of the principal chiral model plus Wess\textendash{}Zumino term,''
\doilink{J. Phys. A \textbf{53} (2020) no.50, 505401}{
doi:10.1088/1751-8121/abc43d}
\arxivlink{2009.00341}.


\bibitem{klim} 
C.~Klimcik,
``Yang-Baxter sigma models and dS/AdS T-duality,''
\doilink{JHEP \textbf{12}, 051 (2002)}{doi:10.1088/1126-6708/2002/12/051}
\arxivlink{hep-th/0210095};
``On integrability of the Yang-Baxter sigma-model,''
\doilink{J. Math. Phys. \textbf{50}, 043508 (2009)}{doi:10.1063/1.3116242}
\arxivlink{0802.3518}.


\bibitem{SST}
K.~Sfetsos, K.~Siampos and D.~C.~Thompson,
``Generalised integrable \ensuremath{\lambda} - and \ensuremath{\eta}-deformations and their relation,''
\doilink{Nucl. Phys. B \textbf{899} (2015), 489-512}{
doi:10.1016/j.nuclphysb.2015.08.015}
\arxivlink{1506.05784}.


\bibitem{lambda} 
K.~Sfetsos,
``Integrable interpolations: From exact CFTs to non-Abelian T-duals,''
\doilink{Nucl.\ Phys.\ B {\bf 880}, 225 (2014)}{doi:10.1016/j.nuclphysb.2014.01.004}
\arxivlink{1312.4560}.


\bibitem{CLa}
%
K.~Sfetsos and K.~Siampos,
``Gauged WZW-type theories and the all-loop anisotropic non-Abelian Thirring model,''
\doilink{Nucl. Phys. B \textbf{885} (2014), 583-599}
{doi:10.1016/j.nuclphysb.2014.06.012}
\arxivlink{1405.7803}.



\bibitem{ZM} V.~E.~Zakharov and A.~V.~Mikhailov,
``Relativistically Invariant Two-Dimensional Models in Field Theory Integrable by the Inverse Problem Technique'',
Sov.\ Phys.\ JETP {\bf 47}, 1017 (1978)
[Zh.\ Eksp.\ Teor.\ Fiz.\ {\bf 74}, 1953 (1978)].

\bibitem{Lund} F.~Lund and T.~Regge,
``Unified Approach to Strings and Vortices with Soliton Solutions,''
\doilink{Phys.\ Rev.\ D {\bf 14}, 1524 (1976)}{doi:10.1103/PhysRevD.14.1524}.


\bibitem{Nappi}
C.~R.~Nappi,
``Some properties of an analog of the nonlinear $\sigma$ model,''
\doilink{Phys.\ Rev.\ D {\bf 21}, 418 (1980)}{doi:10.1103/PhysRevD.21.418}.



\bibitem{GS}
R.~R.~Metsaev and A.~A.~Tseytlin,
``Type IIB superstring action in AdS(5) x S**5 background,''
\doilink{Nucl. Phys. B \textbf{533} (1998), 109-126}{
doi:10.1016/S0550-3213(98)00570-7}
\arxivlink{hep-th/9805028}.

I.~Bena, J.~Polchinski and R.~Roiban,
``Hidden symmetries of the AdS(5) x S**5 superstring,''
\doilink{Phys. Rev. D \textbf{69} (2004), 046002}{
doi:10.1103/PhysRevD.69.046002}
\arxivlink{hep-th/0305116}.

\bibitem{osten}
D.~Osten,
``Lax pairs for new $\mathbb{Z}_N$-symmetric coset \ensuremath{\sigma}-models and their Yang-Baxter deformations,''
\doilink{Nucl. Phys. B \textbf{981} (2022), 115856}{
doi:10.1016/j.nuclphysb.2022.115856}
\arxivlink{2112.07438}.

\bibitem{BR}
H.~A.~Ben\'\i{}tez and V.~O.~Rivelles,
``Yang-Baxter deformations of the $AdS_{5}\times S^{5}$ pure spinor superstring,''
\doilink{JHEP \textbf{02} (2019), 056}{
doi:10.1007/JHEP02(2019)056}
\arxivlink{1807.10432}.

\bibitem{BS}
H.~A.~Ben\'\i{}tez and D.~M.~Schmidtt,
``$\lambda$-deformation of the $AdS_{5}\times S^{5}$ pure spinor superstring,''
\doilink{JHEP \textbf{10} (2019), 108}{
doi:10.1007/JHEP10(2019)108}
\arxivlink{1907.13197}.

\bibitem{DMV0}
F.~Delduc, M.~Magro and B.~Vicedo,
``On classical $q$-deformations of integrable sigma-models,''
\doilink{JHEP \textbf{11} (2013), 192}{
doi:10.1007/JHEP11(2013)192}
\arxivlink{1308.3581}.

\bibitem{HMS0}
T.~J.~Hollowood, J.~L.~Miramontes and D.~M.~Schmidtt,
``Integrable Deformations of Strings on Symmetric Spaces,''
\doilink{JHEP \textbf{11}, 009 (2014)}{doi:10.1007/JHEP11(2014)009}
\arxivlink{1407.2840}.



\bibitem{KY}
D.~Kagan and C.~A.~S.~Young,
``Conformal sigma-models on supercoset targets,''
\doilink{Nucl. Phys. B \textbf{745} (2006), 109-122}{
doi:10.1016/j.nuclphysb.2006.02.027}
\arxivlink{hep-th/0512250}.



\bibitem{eta}
F.~Delduc, M.~Magro and B.~Vicedo,
``An integrable deformation of the $AdS_5 \times S^5$ superstring action,''
\doilink{Phys. Rev. Lett. \textbf{112} (2014) no.5, 051601}{
doi:10.1103/PhysRevLett.112.051601}.
\arxivlink{1309.5850}.


\bibitem{Arut}
G.~Arutyunov, S.~Frolov, B.~Hoare, R.~Roiban and A.~A.~Tseytlin,
``Scale invariance of the $\eta$-deformed $AdS_5\times S^5$ superstring, T-duality and modified type II equations,''
\doilink{Nucl. Phys. B \textbf{903} (2016), 262-303}{
doi:10.1016/j.nuclphysb.2015.12.012}
\arxivlink{1511.05795}.

\bibitem{HMS}
T.~J.~Hollowood, J.~L.~Miramontes and D.~M.~Schmidtt,
``An Integrable Deformation of the $AdS_5 \times S^5$ Superstring,''
\doilink{J. Phys. A \textbf{47} (2014) no.49, 495402}{
doi:10.1088/1751-8113/47/49/495402}
\arxivlink{1409.1538}.


\bibitem{HLS}
B.~Hoare, N.~Levine and F.~K.~Seibold,
``Bi-$\eta$ and bi-$\lambda$ deformations of $\mathbb{Z}_4$ permutation supercosets,''
\arxivlink{2212.08625}.



\bibitem{BEM}
F.~Delduc, B.~Hoare, T.~Kameyama, S.~Lacroix and M.~Magro,
``Three-parameter integrable deformation of $\mathbb{Z}_4$ permutation supercosets,''
\doilink{JHEP \textbf{01} (2019), 109}{
doi:10.1007/JHEP01(2019)109}
\arxivlink{1811.00453}.

F.~K.~Seibold,
``Two-parameter integrable deformations of the $AdS_3 \times S^3 \times T^4$ superstring,''
\doilink{JHEP \textbf{10} (2019), 049}
{doi:10.1007/JHEP10(2019)049}
\arxivlink{1907.05430}.



\bibitem{BL}
C.~Bassi and S.~Lacroix,
``Integrable deformations of coupled $\sigma$-models,''
\doilink{JHEP \textbf{05} (2020), 059}{
doi:10.1007/JHEP05(2020)059}
\arxivlink{1912.06157}.



\bibitem{GSf}
G.~Georgiou and K.~Sfetsos,
``The most general $\lambda$-deformation of CFTs and integrability,''
\doilink{JHEP \textbf{03} (2019), 094}{
doi:10.1007/JHEP03(2019)094}
\arxivlink{1812.04033}.

\bibitem{Falk}
F.~Hassler,
``RG flow of integrable $\mathcal{E}$-models,''
\arxivlink{2012.10451}.


\bibitem{Vdi}
B.~Vicedo,
``On integrable field theories as dihedral affine Gaudin models,''
\doilink{Int. Math. Res. Not. \textbf{2020} (2020) no.15, 4513-4601}{
doi:10.1093/imrn/rny128}
\arxivlink{1701.04856}.

\bibitem{unif}
F.~Delduc, S.~Lacroix, M.~Magro and B.~Vicedo,
``A unifying 2d action for integrable $\sigma$-models from 4d Chern-Simons theory,''
\doilink{Lett. Math. Phys. \textbf{110} (2020), 1645-1687}{
doi:10.1007/s11005-020-01268-y}
\arxivlink{1909.13824}.


\bibitem{HLT} 
B.~Hoare, N.~Levine and A.~A.~Tseytlin,
``Integrable 2d sigma models: quantum corrections to geometry from RG flow,''
\doilink{Nucl. Phys. B \textbf{949}, 114798 (2019)}{10.1016/j.nuclphysb.2019.114798}
\arxivlink{1907.04737}; 
``Integrable sigma models and 2-loop RG flow,''
\doilink{JHEP {\bf 1912}, 146 (2019)}{doi:10.1007/JHEP12(2019)146}
\arxivlink{1910.00397}.

\bibitem{BW}
R.~Borsato and L.~Wulff,
``Target space supergeometry of $\eta$ and $\lambda$-deformed strings,''
\doilink{JHEP \textbf{10} (2016), 045}{
doi:10.1007/JHEP10(2016)045}
\arxivlink{1608.03570}.


\bibitem{unimod}
B.~Hoare and F.~K.~Seibold,
``Supergravity backgrounds of the $\eta$-deformed AdS$_2 \times S^2 \times T^6 $ and AdS$_5 \times S^5$ superstrings,''
\doilink{JHEP \textbf{01} (2019), 125}{
doi:10.1007/JHEP01(2019)125}
\arxivlink{1811.07841}.

S.~J.~van Tongeren,
``Unimodular jordanian deformations of integrable superstrings,''
\doilink{SciPost Phys. \textbf{7} (2019), 011}{
doi:10.21468/SciPostPhys.7.1.011}
\arxivlink{1904.08892}.

\bibitem{PSC}
N.~Berkovits and P.~S.~Howe,
``Ten-dimensional supergravity constraints from the pure spinor formalism for the superstring,''
\doilink{Nucl. Phys. B \textbf{635} (2002), 75-105}{
doi:10.1016/S0550-3213(02)00352-8}
\arxivlink{hep-th/0112160}.

\bibitem{CWY} 
K.~Costello, E.~Witten and M.~Yamazaki,
``Gauge Theory and Integrability, I,''
\doilink{ICCM Not. 6, 46-191 (2018)}{10.4310/ICCM.2018.v6.n1.a6}
\arxivlink{1709.09993}.

K.~Costello and M.~Yamazaki,
``Gauge Theory And Integrability, III''
\arxivlink{1908.02289}.

\end{thebibliography}
\end{document}